\documentclass[useAMS]{mn2e_modmargin}
\usepackage{amsmath}
\usepackage{url}
\usepackage{amsfonts}
\usepackage{amsbsy}
\usepackage{subfigure}
\usepackage{verbatim}
\usepackage{amssymb}
\usepackage{amsbsy}
\usepackage{graphicx}

\newcommand{\omi}{\ensuremath{\boldsymbol{\omega}}}

\newcommand{\Ro}{\ensuremath{\text{Ro}}}

\newcommand{\mP}{\ensuremath{\overline{P}}}
\newcommand{\mrho}{\ensuremath{\overline{\rho}}}

\newcommand{\ou}{\ensuremath{\overline{\mathbf{u}}}}

\renewcommand{\u}{\ensuremath{\mathbf{u}}}
\renewcommand{\d}{\ensuremath{\partial}}

\newcommand{\ex}{\ensuremath{\mathbf{e}_{x}}}
\newcommand{\ey}{\ensuremath{\mathbf{e}_{y}}}
\newcommand{\ez}{\ensuremath{\mathbf{e}_{z}}}
\newcommand{\ephi}{\ensuremath{\mathbf{e}_{\phi}}}

\title[Local models of discs]
 {Local models of astrophysical discs}
\author[H. N. Latter \& J. Papaloizou]{Henrik N. Latter\thanks{E-mail:
    hl278@cam.ac.uk} \& John Papaloizou \\
 DAMTP, University of Cambridge, CMS, Wilberforce Road,
Cambridge CB3 0WA, UK}

\begin{document}

\maketitle

\begin{abstract}

Local models of gaseous accretion discs have been successfully
employed for decades to describe an assortment of small scale phenomena,
from instabilities and turbulence, to dust dynamics and planet formation.
For the most part, they have been derived in a physically motivated
but essentially ad hoc fashion, with some of the mathematical
assumptions never made explicit nor checked for consistency.
This approach is 
susceptible to error, and it is easy to derive local models that
support spurious instabilities or fail
to conserve key quantities. In this paper we 
present rigorous derivations, based on an asympototic 
ordering, and formulate a hierarchy of local models
(incompressible, Boussinesq, and compressible), making clear which is
best suited for a particular flow or phenomenon, while spelling out
explicitly the assumptions and approximations of each. 
We also discuss the merits of the anelastic approximation, emphasising 
that anelastic systems struggle to conserve energy
unless strong restrictions are imposed on the flow.
The problems encountered by the anelastic approximation are
exacerbated by the disk's differential rotation, but also attend
non-rotating systems such as stellar interiors.
We conclude with a defence of local models and
their continued utility in astrophysical research.

\end{abstract}


\section{Introduction}

Astrophysical discs exhibit dynamical phenomena on a wide range of scales, from the global
(e.g.\ warps, outbursts,
spiral waves, and outflows) to the small-scale (turbulence,
planet formation, etc). In order to attack
the latter it is convenient, analytically
and computationally, to deploy a local model: i.e. to `zoom in' on a
small patch of disc, approximate it as Cartesian, and treat it
in isolation of the rest of the system. This technique is commonly employed
 in the study of
celestial mechanics (cf.\ the Hill equations), galactic dynamics
(Goldreich and Lynden-Bell 1965), planetary rings (Wisdom and Tremaine
1988), and gaseous accretion discs (made
especially famous by application to the magnetorotational instability,
MRI; Hawley et al.~1995). 

The derivation of local models for particulate discs is relatively
straightforward, but there are subtleties involved when dealing with a
gas, and its thermodynamic variables. 
Depending on the properties of the flow, in particular the sizes
of its
characteristic lengthscales and Mach number, certain terms in the
governing equations are dominant, some
negligible; as a consequence, we are led to a number of different
local approximations: 
incompressible,
Boussinesq, anelastic, and small and large compressible boxes. 
Though regularly used,
their derivations have been ad hoc for the most part, 
and though physically motivated it is easy to derive equations that
fail to conserve key properties or, even worse, introduce spurious
instabilities. These problems arise especially when attempting to
incorporate the background thermodynamic gradients, 
and can be connected to the violation of wave-action conservation. 

The main aim of this paper is to highlight these issues while
rigorously deriving
local approximations that are unambiguously consistent and conservative, hence fit for
purpose. The essential assumptions of each model are clearly spelled
out so that each may be matched to the appropriate problem.
They may then serve as a set of references for researchers in
the field.
We employ an ordering
approach similar in some details to Spiegel and Veronis (1960) and Gough (1969). The
different models can then be clearly delineated in terms of a handful
of key dimensionless parameters, such as $\lambda/R$, $\lambda/H_Z$,
and $\mathcal{M}$, where $\lambda$ is a characteristic length scale of the
flow, $R$ is disk radius, $H_Z$ is the vertical
pressure scale height, and $\mathcal{M}$ is the Mach number of the
perturbed flow in the corotating frame. 

Starting from a flow in a fully global and fully compressible disc,
we first derive the equations governing
the incompressible shearing box, the independent 
distinguishing features of
which are (a) the flow is small-scale, with a characteristic
lengthscale much less than the scale height, $\lambda/H_Z\ll 1$, 
(b) it is subsonic $\mathcal{M}\ll 1$, 
and (c) the fractional density and
pressure perturbations are equally small (note, however, that the
density plays no part in the final equations). The Boussinesq box,
which we deal with next, also
 assumes (a) and (b) but instead of (c) lets the fractional density
perturbation take significantly larger values than the
fractional
pressure perturbation (though both must still be $\ll 1$). The last
move permits the
inclusion of a constant background entropy gradient, and hence
buoyancy effects.  
We show how weak vertical
shear may be incorporated consistently, and how potential vorticity
manifests in these equations.

Anelastic models generally assume slow flows and
small
thermodynamic perturbations, as earlier, but permit the flow to range
over longer length-scales, up to the scale height, i.e.\ 
$\lambda/H_Z\sim 1$. We highlight the inability
of this approximation to conserve energy under general conditions, 
and we derive a consistent set of conservative
equations that resemble the so-called `pseudo-incompressible'
approximation (Durran 1989, Vasil et al.~2013),
 but which requires additional restrictions on the lengthscale of the
 flow. These additional restrictions issue directly from the strong
 differential rotation and illustrate
 the challenges in combining anelasticity and astrophysical disks.
Note that even in the absence of strong rotation and shear, as in the
stellar context, the anelastic
equations remain extremely problematic (see Brown et al.~2012).

Compressible
models impose no restriction on the flow speed, 
and hence the Mach number $\mathcal{M}$ can take any value. We obtain
the `small compressible box' by assuming small scales 
$\lambda/H_Z\ll 1$, a model that
cannot involve any background thermodynamic gradients, 
and the `large compressible box', by
assuming $\lambda/H_Z\sim 1$, which must include the full vertical
structure of the disc. We devote time to exposing the spurious
instabilities that appear when the background 
gradients are included incorrectly, and discuss how they
are associated with a breakdown of wave-action conservation.

The paper concludes with a short defence of local models
and their importance for research in astrophysical accretion
flows. We argue that previous criticisms are overstated, and that
global set-ups suffer from problems of a similar or greater
magnitude. Our essential point is that, owing to the simpler geometry afforded by local models,
researchers can more easily disentangle the salient physical processes
and their interconnection, and as a consequence develop a deeper understanding of
the overall astrophysical flow. Local model help us work out the
conceptual
ideas and physical intuition necessary for the interpretation
of observations and global simulations, which are usually 
much more complicated and messy. 

The paper is lengthy and technically detailed. In order to ease its
readability,
each of Sections 3 to
6 are as self-contained as possible and may be read independently. 
All four section, however, make use of the material in Section 2, in which
we present our background
global disk (with its key length and timescales) and
introduce the thin disk and local approximations, both necessary to
derive the equations of any shearing box. 
Section 3 presents a derivation of the incompressible shearing
box, and Section 4 the Boussinesq box, both from this global, compressible
starting point. In Section 5, we discuss
anelastic models and their problems, while developing a consistent
and conservative set of
anelastic equations.  In
Section 6, we present compressible models, purely local (`small box') and vertically
stratified (`large box'), emphasising the spurious instabilities that arise if the
equations are incorrectly derived. 
Our conclusions appear in Section 7. 
For readers who wish to skip the derivations and go straight 
to the presentation of the final
equations associated with each model, the incompressible shearing box equations appear in Section
3.3, the Boussinesq equations in Section 4.3, the consistent anelastic
equations in Section 5.2.4, and the compressible equations in Sections
6.1.1 and 6.2.

\section{Preliminaries}

To begin, we display the equations governing fluid flow 
in a representative disc
alongside its basic global equilibrium. 
It is upon this quasi-steady equilibrium state that
fast small-scale phenomenon (captured by local models) manifests. 
We subsequently introduce the additional
assumptions of a thin disc and
small-scales.

\subsection{Governing equations}

Consider an inviscid astrophysical disc, rotating about a central object.
Let us describe it in a rotating cylindrical reference frame
with rotation vector $\boldsymbol{\Omega}=\Omega_0\,\ez$, where
$\Omega_0$ is a constant frequency specified later. We assume that the disk is
radially extended, and thus do not consider slender tori or narrow
rings, nor their local representations (see for e.g.\ Goldreich et
al.~1986 and Narayan et al.~1987)
The equations controlling the flow are 
\begin{align}
\frac{D\rho}{Dt}  &= -\rho\nabla\cdot \u, \label{i1}\\
\frac{D\u}{Dt} &= -2\Omega_0\ez\times\u 
  -\frac{1}{\rho}\nabla P - \nabla\Phi, \label{i2}\\
\frac{DS}{Dt}  &= \Gamma - \Lambda,\label{i3}
\end{align}
where $\rho$ is volumetric mass density, $\u$ is velocity, and $P$ is
pressure. The symbol $\Phi$
combines the centrifugal potential $-\frac{1}{2}\Omega_0 R^2$, with
$R$ cylindrical radius, and the
gravitational potential of the star $\Phi_*$. To make life
simple we assume that the star possesses a point mass potential.
In addition, $S$ denotes the entropy function, 
$\Gamma$ represents external heating (from
the star or cosmic rays, perhaps), and $\Lambda$ represents
radiative cooling (which may take the form $-\nabla\cdot\mathbf{F}$,
where $\mathbf{F}$ is a radiative flux). It is assumed that $\Gamma$
and $\Lambda$ are functions of the spatial coordinates,
or else of the thermodynamic variables. In what follows, often we
combine both cooling and heating into the single function $\Xi$.
Finally, the disc is composed of an ideal gas, so
that $P=\mathcal{R}\rho T$, where $\mathcal{R}$ is the gas
constant, and $T$ is temperature.

\subsection{Equilibrium}

A local model's small-scale phenomena take place upon a
pre-existing equilibrium state.
An assumption,
shared by all local models, is that the smaller-scale processes \emph{do not
feed back} onto the background equilibrium. The equilibrium
remains fixed on the short timescales of interest.

We assume that the disc falls into the steady state
\begin{align}
\u=\overline{\u} \equiv R\Omega(R,Z)\,\ephi, \quad P=\overline{P}(R,Z), \quad \rho=\overline{\rho}(R,Z),\label{eqvel}
\end{align}
where the three functions $\Omega$, $\mP$, and $\mrho$ may be obtained from the
equations:
\begin{align}
&\ou\cdot\nabla\ou = -2\Omega_0 \ez\times \ou -\frac{1}{\mrho}\nabla\mP - \nabla\Phi, \label{eqm1} \\
&\Gamma = \Lambda. \label{eqm2}
\end{align}
Note that $\Omega$ is the orbital frequency of the gas in the
rotating frame. 
Because $\Omega$ does not appear in the energy balance, we can
obtain $\mrho$ and $\mP$ from the vertical component of
Eq.~\eqref{eqm1} and from Eq.~\eqref{eqm2}. Then
the radial component of \eqref{eqm1} obtains $\Omega$. 
For reference, Eqs \eqref{eqm1} may be recast as
\begin{align} 
&-R(\Omega+\Omega_0)^2 =
-\frac{1}{\mrho}\d_R\mP-\d_R\Phi,\label{eqm3}\\
& \frac{1}{\mrho}\d_Z\mP=-\d_Z\Phi.\label{eqm4}
\end{align}
We next define three fundamental lengthscales. 
The equilibrium is taken to vary relatively smoothly with respect to the spatial
coordinates and that $\mrho$, $\mP$, and $\Omega$ exhibit well-defined
scales of variation. We introduce the radial pressure scale
height $H_R$, and the vertical pressure scale height $H_Z$. These
possess the scalings
\begin{align}
&H_R^{-1} \sim \d_R \ln \mP \sim \d_R \ln \mrho, \\
&H_Z^{-1} \sim \d_Z \ln \mP \sim \d_Z \ln \mrho. 
\end{align}
Note that it has been assumed that the density and pressure scale heights are
of the same order of magnitude, which is reasonable in most contexts. 

An astrophysical disc, perhaps by definition, is rotationally
supported, 
so the scale
of $\Omega$'s radial
variation will be $\sim R$. Its vertical scale of variation we denote
by $H_\Omega$, and it satisfies
\begin{equation}
H_\Omega^{-1} \sim \d_Z \text{ln}\Omega.
\end{equation}
In fact, this length can be estimated from the thermal wind equation,
as shown in the next subsection. 

\subsection{The local and thin-disc approximations}

The equilibrium is disturbed
so that $\rho=\mrho+\rho'$, etc, where a prime denotes a
perturbation. 
The (nonlinear) perturbation equations are
\begin{align}
\frac{D \rho'}{Dt} &= -(\mrho+\rho')\nabla\cdot \u'-\u'\cdot\nabla\mrho, \label{nl1}\\
\frac{D \u'}{Dt} &= -\frac{1}{\mrho+\rho'}\nabla P' + \frac{\nabla
  \mP}{\mrho(\mrho+\rho')}\,\rho'\notag\\ & \hskip1cm  -2\Omega_0
\ez\times\u'-\u'\cdot\nabla \ou, \label{nl2} \\
 \frac{D S'}{Dt} &= \Gamma-\Lambda - \u'\cdot\nabla \overline{S} \label{nl3}
\end{align}
where now $D/Dt= \d_t + (\ou+\u')\cdot\nabla$, and $\overline{S}$ is the
equilibrium entropy distribution.

The perturbations
exhibit a characteristic lengthscale $\lambda$. We could in fact
specify lengthscales in all three directions
$\lambda_R$, $\lambda_\phi$, and $\lambda_Z$, but it suffices to 
designate only one length, at this stage. 
We also assume the phenomena exhibits a characteristic
velocity scale $w$, and thus a characteristic timescale of $\lambda/w$.
The first essential assumption
that we make, and which is shared by all local models, is that at any
radial location
$\lambda \ll R$. The perturbations are small scale relative to radius.

The next step is to exclusively focus upon a specific
location in the disc
\begin{equation}
R=R_0, \quad \phi=\phi_0, \quad Z=Z_0.
\end{equation}
Then we \emph{choose}
the rotation frequency of our frame of reference $\Omega_0$ 
so that $\Omega(R_0,\,Z_0)=0$. This means that at this location the disc's
orbital frequency is equal to the frame's rotation rate.
We introduce spatial variables centred upon this location
\begin{align}
x= R-R_0,\qquad y=R_0(\phi-\phi_0),\qquad z= Z-Z_0,
\end{align}
and suppose that they take values over a range of order
$\lambda$. Hence $x,y,z \ll R_0$. For the moment we do not specify the
relative sizes of $\lambda$ and $Z_0$. 
It is not hard to see that,
to leading order in $\lambda/R_0$, the del operator simplifies to
\begin{equation}
 \nabla \approx \ex\d_x + \ey \d_y + \ez \d_z,
\end{equation}
where the new locally Cartesian coordinates $x,y,z$ point in the
radial, azimuthal, and vertical directions. All the cylindrical
terms are subdominant, because they are $\sim \lambda/R_0 \ll 1$ smaller.

The second essential assumption is that the disc is
thin. Mathematically, this corresponds to $H_Z \ll R_0$. 
Because our local box
must be located in the bulk of the disc, we assume that
$Z_0 \lesssim H_Z$. Furthermore, the new variable $z$ cannot be much
greater than $H_Z$. The assumption of a thin disc is crucial as it
permits us to obtain a dimensional estimate on the equilibrium pressure.
From Eq.~\eqref{eqm4}, expanding the point mass potential
$\Phi$ in small $Z/R$ yields
\begin{align*}
\d_Z \mP = \mrho\Omega_0^2 Z.
\end{align*}
Assuming $Z\sim H_Z$ and $\d_Z \sim 1/H_Z$ gives the following scaling
for the equilibrium pressure
\begin{equation} \label{pressure}
\mP \sim \mrho H_Z^2\Omega_0^2,
\end{equation}
one that is unique to astrophysical discs. We then recognise that the sound speed of
the gas $c$ scales as $H_Z\Omega_0$.

Finally, we take the curl of \eqref{eqm1} and write its $\phi$-component
as
\begin{equation} \label{tw}
R\d_Z\Omega^2 = \frac{1}{\mrho^2}\left(\nabla\mP\times\nabla \mrho\right)\cdot\ephi.
\end{equation}
This is the `thermal wind equation'. It helps us assess the amount
of vertical shear in the equilibrium. Only baroclinic
equilibria, in which pressure depends on both density and another
thermodynamic variable, permit a nonzero $\d_Z\Omega$. Using \eqref{pressure} and
the characteristic lengthscales introduced in the previous subsection,
we obtain the following estimate:
\begin{equation}
H_\Omega \sim \left(\frac{H_R}{H_Z}\right)R_0 > R_0.
\end{equation}
Thus the characteristic vertical lengthscale of the vertical shear is
long, greater than $R_0$ in fact.

In summary, we have introduced two assumptions, shared by all shearing
box models. First: radial locality, i.e.\ that
$\lambda \ll R_0$. Second: the thin disc approximation, i.e.\ $H_Z \ll
R_0$. These assumptions lead to a sequence of local models, their
distinguishing features resting on the size of (a) the ratio of lengths
\begin{equation}
\epsilon \equiv \lambda/H_Z,
\end{equation}
(b)  the
Mach number of the flow
\begin{equation}
\mathcal{M} \equiv \frac{w}{H_Z\Omega_0},
\end{equation}
and (c) the size of the fractional perturbations in
pressure and density. In Table I we list the most important
parameters, scales, and definitions that appear in the following derivations.

\begin{table}
\begin{tabular}{|c|c|}
 \hline
 Symbols & Definitions \\
 \hline
$\mrho$, $\ou$, $\mP$, $\overline{S}$ & Equilibrium quantities \\
$H_R$, $H_Z$ & Eq'm radial and vertical scaleheights \\
$H_\Omega$  & Lengthscale of eq'm vertical shear \\
$q_R=(\d\ln\Omega/\d\ln R)_0$ & Dimensionless radial shear rate \\
 $q_Z=(R\d\ln\Omega/\d Z)_0$, & Dimensionless vertical shear rate \\
$\Phi_T=\Omega_0^2 q_R x^2,$  & Radial part of the tidal
potential \\
$\Phi_Z=\tfrac{1}{2}\Omega_0^2 z^2,$  & Vertical part of the tidal potential \\
$\rho'$, $\u'$, $P'$, $S'$ & Perturbations \\
$\lambda$, $w$ & Perturbation length and velocity scales \\
$\rho^*$, $\u^*$, $P^*$, $S^*$ & Dimensionless perturbations \\
$\boldsymbol{\xi}$& Lagrangian displacement \\
$\epsilon=\lambda/H_z$ & Lengthscale ratio \\
$\mathcal{M}= w/(H_Z\Omega_0)$ & Mach number \\
$ \Ro = w/(\lambda\Omega_0)$  & Rossby number \\
$\ell=u_y + 2\Omega_0 x$ & Specific angular momentum  \\
$\omi= \nabla\times \u + 2\Omega_0 \ez$ & Vorticity \\
$\Theta$    & Potential vorticity \\
 \hline
\end{tabular}
\caption{Important symbols, characteristic scales, and definitions.}
\end{table}

\section{Incompressible dynamics}\label{Incompress}

\subsection{Distinguishing assumptions}

We begin with the derivation of a purely incompressible
model, suitable for slow dynamics: shear instability, convection (both
radial and vertical), vortices, and the MRI. 
First, we assume that $\lambda \ll H_Z,\, H_R$, which means the
parameter $\epsilon\ll 1$. 
If $Z_0 \lesssim H_Z$, we can expand
the equilibrium function $\Omega$ in Eqs \eqref{nl1}-\eqref{nl2} in small
$x$ and $z$ and only retain the leading order terms. This move is
acceptable because
$\Omega$'s characteristic lengthscales of variation are larger than
that of the phenomena in question. 
In fact,
\begin{equation}
 \Omega(R_0+x,\,Z_0+z) = (\d_R\Omega)_0 x + (\d_Z\Omega)_0 z +
 \mathcal{O}(x^2,z^2),
\end{equation}
where the subscript
the $0$ indicates evaluation at the centre of the box.
Note that the leading order constant term in the expansion of $\Omega$ is zero. 
Moreover, if we locate the shearing box at the midplane, $Z_0=0$, then
$(\d_Z\Omega)_0=0$ because of symmetry: in this special case there
is no vertical shear in the box. In general, however, we
write
\begin{equation}
\ou = \Omega_0(q_R\,x + q_Z \,z)\ey,\label{ueq}
\end{equation}
where
\begin{equation} \label{queues}
q_R \equiv (\d\ln\Omega/\d\ln R)_0\quad \text{and} \quad q_Z \equiv R_0(\d\ln\Omega/\d
Z)_0. 
\end{equation}
It is worth noting at this point that while typically $q_R\sim
1$, we have $q_Z \sim H_Z/H_R$ and so could be considerably smaller
(though not necessarily as small as $\epsilon$) and
possible to omit. If $Z_0=0$, then $q_Z=0$ exactly.

What of the equilibrium 
thermodynamic variables $\mP$ and $\mrho$? Because we
are only interested in very short scales $\lambda$, much smaller than
the variation in $\mP$ and $\mrho$, we do not expect $\mP$ and $\mrho$
to change greatly in our box. In particular, $\mrho \approx \mrho_0$,
true to leading order in $\epsilon$, where $\mrho_0=\mrho(R_0,Z_0)$, a
constant. Similarly $\nabla\mP \approx (\nabla \mP)_0$, a constant
vector. 

The second assumption we make concerns the Mach number of the flow. 
`Slow' approximations,
such as the incompressible, Boussinesq, and anelastic models, set
$\mathcal{M}\ll 1$, and are valuable because they filter out sound waves
that may pose numerical challenges and prevent analytical progress. 
We thus have
two small ordering parameters $\epsilon$ and $\mathcal{M}$, which are
instructive to keep separate (though it is possible to equate them). 
In fact, if we set $\mathcal{M}\sim \epsilon$ then we are unduly
restricting the kinds of phenomena described, ensuring their
characteristic timescale is $1/\Omega$, and thus pinned to the orbital period. 

The third and final assumption deals with the sizes of the thermodynamic perturbations
$\rho'$ and $P'$. They must remain very small compared to the
background but stay the same magnitude as each other. We let
\begin{align}
\frac{\rho'}{\mrho} \sim \frac{P'}{\mP} \sim \mathcal{M}^2.
\end{align}
It might seem strange to force
the pressure perturbations to be small in a nominally incompressible
model 
 --- but what matters most are the
pressure gradients, and because the spatial scales of variation are so
small, $\lambda\ll H_Z$, the gradients are potentially huge. The
pressure perturbation must be scaled appropriately so the pressure term
does not blow up.

\subsection{Derivation}

We are now in a position to derive our equations, by
rescaling all the variables and collecting
the leading order terms in $\epsilon$ 
and $\mathcal{M}$. The
perturbations may be written as
\begin{align}\label{perturbed}
\u' = w \u^*, \quad \rho'= \mathcal{M}^2 \mrho_0 \rho^*,
\quad P' = \mathcal{M}^2 \mrho_0 H_Z^2\Omega_0^2 \, P^*,
\end{align}
where the star indicates an order 1 dimensionless variable. The
spatial variables are $x = \lambda x^*$, etc, and time is
$t=t^*(w/\lambda)$. Acknowledging the scaling Eq.~\eqref{pressure}, the
background equilibrium may be non-dimensionalised as
\begin{align}\label{eqmscale1}
&(\d_Z \mP)_0 = \mrho_0 H_Z\Omega_0^2 (\d_Z \mP)_0^*, \quad 
(\d_R \mP)_0 = \mrho_0 \frac{H_Z^2}{H_R}\Omega_0^2 (\d_R \mP)_0^*, \\
&(\d_Z \mrho)_0 = \mrho_0 \frac{1}{H_Z} (\d_Z \mrho)_0^*, \quad 
(\d_R \mrho)_0 = \mrho_0 \frac{1}{H_R} (\d_R \mrho)_0^*,\label{eqmscale2}
\end{align}
and
\begin{equation} \label{youse}
\ou = \lambda\Omega_0\ou^* = \lambda\Omega_0 (q_R\,x^* + q_Z\,z^*)\ey.
\end{equation}

Expressions \eqref{perturbed}-\eqref{youse} are thrown
into the continuity equation
Eq.~\eqref{nl1}. 
There is only a single order 1 term, with respect to both $\epsilon$
and $\mathcal{M}$, and the equation
reduces to the incompressibility condition
\begin{equation}
\nabla^* \cdot\u^* = 0.
\end{equation}
On the other hand, the order 1 terms in the momentum equation are
\begin{align}
\frac{D\u^*}{Dt^*}
 &= -\nabla^*P^* -2\Ro^{-1}\ez\times\u^* \notag\\
 &\hskip2.5cm - \Ro^{-1}(q_R\,u_x^* +q_Z\,u_z^*)
\ey, \label{momincompressible}
\end{align}
where $D/Dt^* = \d_t^*+ (\u^*+\Ro^{-1}\ou^*)\cdot\nabla$.
The background pressure gradient is $\epsilon$ smaller than the
terms above and is hence dropped. Also, the Rossby number
\begin{equation}
\Ro \equiv \frac{w}{\lambda\Omega_0}
\end{equation}
appears,
which quantifies the importance of the differential rotation. When the
characteristic frequency of the phenomenon exceeds $\Omega_0$ then
$\Ro$ increases and the Coriolis and shear terms become
subdominant. In fact, in this limit
the shearing box is isotropic and homogeneous.  
This regime manifests on sufficiently short scales if $w$ does not increase
concomitantly with $\lambda$. Thus it is to be expected that a typical hydrodynamical
turbulent cascade 
ultimately reaches $\Ro\gg 1$ on a small enough scale, flow on these
short scales ignorant of shear and rotation. 
In some applications 
(e.g. involving well-coupled dust) it may not be
necessary to describe the system with a shearing
box at all. 

\subsection{Final incompressible equations}

In dimensional form the final
system is
\begin{align}
\frac{D\u'}{Dt} &= -\frac{1}{\mrho_0}\nabla P' -2\Omega_0\ez\times\u' - 
                 \Omega_0(q_R\,u_x +q_Z\,u_z) \ey, \\
\nabla \cdot \u' &=0,
\end{align}
where $D/Dt = \d_t + \Omega_0 (q_R x + q_Z z)\d_y + \u'\cdot\nabla$.
There is no need for the equation of state, nor the entropy
equation.

In terms of total, rather than perturbed, variables, the
momentum equation may be written as
\begin{equation} \label{inc_mom}
\frac{D\u}{Dt} = -\frac{1}{\mrho_0}\nabla P -2\Omega_0\ez\times\u - 
                 2\Omega_0^2(xq_R  +zq_Z ) \ex,
\end{equation}
where the last term is the tidal force. Note that if $q_Z\neq 0$ then
this tidal force is not conservative (i.e. cannot be written as the
gradient of a scalar).

In summary: to obtain these equations we have adopted (a) radial
locality, (b) the thin disc approximation, as in all shearing box
models, and then assumed that (c) the characteristic scales
are much smaller than the scale heights, i.e.\ $\lambda \ll
H_Z,\,H_R$, (d) the phenomena are very slow, so that the Mach
number is $\mathcal{M}\ll 1$, and (e)
 the fractional thermodynamic perturbations are both $\sim
 \mathcal{M}^2$. 
Note that the last two
conditions are not enforced by the model: simulations that exhibit
extremely strong flows could be violating these assumptions. Runaway
MRI channel flows in incompressible boxes are an example 
that come to mind (Lesaffre et al.~2009). Given, however, that
formally the soundspeed is infinity, it is difficult to judge a
posteriori whether restriction (d) is violated from the simulation
data itself.

\subsection{Conservation laws}

In order to produce physically acceptable dynamics, the governing
equations must conserve certain quantities. Energy, angular
momentum, and vorticity are the
most important. 

\subsubsection{Kinetic energy}

We derive an equation for the specific kinetic energy by taking the
scalar product of Eq.~\eqref{inc_mom} with $\u$ and making repeated
use of the incompressibility condition. We find
\begin{equation}
\d_t(\tfrac{1}{2}u^2) + 
\nabla\cdot\left[(\tfrac{1}{2}u^2+h +\tfrac{1}{2}\Omega_0^2q_Rx^2)\u
\right]= -\Omega_0^2q_Z z\,u_x,\label{CONE}
\end{equation}
where $h=P/\rho_0$ is the pseudo-enthalpy.
Note the source term on the right hand side:
when vertical shear is included in the model, i.e.\ $q_z\neq 0$, then
energy is \emph{not} conserved in the box.

The source term, however, is physical and
 comes from the rate of doing $PdV$ work in a baroclinic flow.
In a barotropic fluid, energy is conserved on a closed
stream tube. But this need not be true if the fluid were baroclinic: 
energy can be input when material flows from a high density region to a low density
region and back via a different path, even when that flow is incompressible.
 In Appendix A, the origin and form of this term is discussed
in greater detail. 

\subsubsection{Angular momentum}

Next we turn to angular momentum, which in the shearing sheet is
represented by the specific canonical $y$-momentum 
\begin{equation}
\ell \equiv u_y + 2\Omega_0 x.
\end{equation}
Rearranging the $y$-component of \eqref{inc_mom} yields
\begin{equation}\label{ell}
\d_t \ell + \u\cdot\nabla\ell = -\d_y P.
\end{equation}
Thus the angular momentum of a fluid blob can only change due to
azimuthal accelerations from the pressure gradient. It follows
that angular momentum is materially conserved in axisymmetric flow, 
and there can be no
accretion in this case; fluid blobs (or rather rings) cannot exchange
angular momentum because they cannot azimuthally accelerate one
another (also see Stone and Balbus 1996). Finally, if we integrate
Eq.~\eqref{ell} over the box, use the incompressibility condition, and
impose periodic
boundaries in $y$, we observe that the total angular momentum of the system is
constant.

\subsubsection{Vorticity}

Finally, we exhibit the vorticity equation. By taking the
curl of \eqref{inc_mom} and applying standard vector identities,
one obtains
\begin{equation}\label{vote}
\d_t\omi +
\u\cdot\nabla\omi-\omi\cdot\nabla\u= -\Omega_0^2 q_Z \ey,
\end{equation}
where 
\begin{equation}\label{vort}
\omi= \nabla\times\u + 2\Omega_0\ez 
\end{equation}
 is the
vorticity in the shearing box.
Note the constant source term on the right side of
Eq.~\eqref{vote}; it is zero only in the absence of vertical shear.
The source term is result of the
baroclinicity of the flow, and in fact, is the local 
manifestation of the $\nabla\mrho\times\nabla \mP$ term in the thermal
wind balance \eqref{tw}. 

The existence of a constant
injection of vorticity possibly
causes problems in simulations of the flow.
 It comes about essentially because of the local
dynamics' inability to react back on the equilibrium conditions. This might
be reasonable when dealing with the radial shear,
but the vertical shear might get smoothed out effectively on short
times by the Goldreich-Schubert-Fricke instability (Goldreich and
Schubert 1967, Fricke 1968, Nelson et al.~2013) 
unless it is forcibly maintained (by powerful stellar
radiation, for instance; Barker and Latter 2015).

\section{The Boussinesq approximation}\label{Bouss}

\subsection{Distinguishing assumptions}

This class of model differs from the previous incompressible case in its
treatment of the density perturbation. As before, we assume that the
phenomena of interest exhibit lengthscales much less than the scale
heights of the disk,
$\lambda \ll H_Z,\,H_R$ so that $\epsilon \ll 1$. As a result
the background fields  $\ou$, $\mP$ and $\mrho$ and
their derivatives may be expanded in small $x$ and $z$. 
The former is hence linear in these variables, taking the form of Eq.~\eqref{ueq}, while the latter
two are constant. In addition, the flow is presumed to be subsonic, so
that $\mathcal{M}\ll 1$. 
However, we assume that
$\rho'/\mrho$ is much larger than $P'/\mP$, though both remain $\ll
1$. More precisely
\begin{equation}
 \frac{\rho'}{\mrho}\sim \frac{\mathcal{M}^2}{\epsilon}, 
\qquad \frac{P'}{\mP}\sim
\mathcal{M}^2, \label{Booziescalings}
\end{equation}
with the additional requirement that $\mathcal{M}^2 \ll \epsilon$.
We could just set $\mathcal{M}\sim \epsilon$ but this is
unnecessary, and in fact imposes an additional
restriction that is undesirable.

\subsection{Derivation}

As in Section 3.2 we proceed by introducing dimensionless
variables. The equilibrium fields
are expressed as in Eqs \eqref{eqmscale1}-\eqref{youse}, while the independent
variables may be written as $x=\lambda x^*$, etc, and $t=(w/\lambda)t^*$,
where a star indicates a dimensionless order 1 quantity.
The perturbations are
\begin{align}
\u' = w \u^*, \,\,\, \rho'= (\mathcal{M}^2/\epsilon) \mrho_0 \rho^*,
\,\,\, P' = \mathcal{M}^2 \mrho_0 H_Z^2\Omega_0^2 \, P^*.
\end{align}
Throwing these into the continuity equation yields to order 1,
\begin{equation}
\nabla^*\cdot\u^* =0,
\end{equation}
the incompressibility condition again.
The momentum equation is more interesting. We obtain
\begin{align}
&\frac{D\u^*}{Dt^*} = -\nabla^*P^* -2 \Ro^{-1}\ez\times\u^* - \Ro^{-1}(q_R\,u_x^* +q_Z\,u_z^*)
\ey \notag\\ & \hskip2cm + \left[\frac{H_Z}{H_R} \left(\d_R \mP
  \right)^*_0\ex + (\d_Z\mP)_0^* \ez  \right]\rho^*, \label{momBoussinesq}
\end{align}
where Ro$=w/(\lambda\Omega_0)$ is the Rossby number, and
$D/Dt^* = \d_t^*+ (\u^*+\Ro^{-1}\ou^*)\cdot\nabla$.
Note the buoyancy term on the right hand side in square
brackets, absent in the incompressible derivation in Section 3.2. 
It prompts a number of comments.

First, if our sheet is located on the midplane, i.e.\
$Z_0=0$, then $(\d_Z\mP)_0=0$ by symmetry and hence there can be no
vertical buoyancy contribution. 

Second, the radial buoyancy term is multiplied by
a factor $H_Z/H_R$. We have been careful, so far, not to
specify explicitly the relative magnitudes of the two scale heights. If
$H_R\sim R_0$, as we might expect for a very smooth thin disc, then the
radial buoyancy term is $\epsilon$ smaller than the other
terms and should be dropped. If however, the disc exhibits more abrupt
radial structure, so that $H_Z\lesssim H_R$,  then the term should be
retained. Dead-zone
edges, gaps opened by planets, or icelines in protoplanetary discs are
examples of structures that could give rise to the latter. 

Third, the new buoyancy term brings in the variable $\rho^*$, in
addition to $P^*$ and $\u^*$, and so another equation is required.
From the definition of the
entropy function $S\propto\ln(P\rho^{-\gamma})$, we have
\begin{equation}
S^* = -\gamma\frac{\rho'}{\mrho} = -(\mathcal{M}^2/\epsilon)\,\gamma\rho^*,
\end{equation}
to leading order, according to the scalings
\eqref{Booziescalings}. Next, in Eq.~\eqref{nl3} the equilibrium entropy
$\nabla\overline{S}$ is expanded in small $x$ and $z$,
 and is, to leading order, an order 1 constant vector. The dominant
 terms in the entropy
 equation are then
\begin{equation} \label{rhoeqn}
\frac{D\rho^*}{Dt^*} = -\frac{1}{\gamma}\Ro^{-2}\u^*\cdot
\left[\frac{H_Z}{H_R}(\d_R\overline{S})_0\ex + (\d_Z
  \overline{S})_0 \ez \right]
+ \Xi^*(\rho^*),
\end{equation}
where the perturbed non-adiabatic contributions have been packaged into the
linear function/operator $\Xi$. An important case is when the external
heating does not depend on $\rho$ and radiative cooling can be
represented using the diffusion approximation. Then $\Xi(\rho^*)=
\eta\nabla^2\rho^*$, with $\eta$ the thermal diffusivity. An alternative
is a cooling law, such as $\Xi \propto -\rho^*$. Whatever form $\Xi$ takes,
however, it should be linear. Note again the coefficient
$H_Z/H_R$, which may be small, but the term is retained, as
explained earlier.

Finally, the entropy gradient term in the $\rho^*$-equation \eqref{rhoeqn} 
is
multiplied by the inverse squared Rossby number. In a typical
hydrodynamic turbulent cascade,
the flow enters the $\Ro\gg 1$ regime on sufficiently short scales; then,
not only does the differential rotation drop
out of the problem, so does the background entropy gradient. As a
result, the
$\rho^*$ perturbation is controlled by simple advection and the
heating/cooling physics embodied in $\Xi$. Certainly the latter we
expect to return $\rho^*$ to zero and the thermal dimension of the
problem likely drops out entirely. 

\subsection{Final Boussinesq equations}

When we return to dimensional variables it is convenient to
introduce the following notation. Define the radial and vertical
buoyancy frequencies by
\begin{align}
N^2_R \equiv -\frac{1}{\gamma \mrho_0} (\d_R \mP)_0 (\d_R \overline{S})_0,
\quad
N^2_Z \equiv -\frac{1}{\gamma \mrho_0} (\d_Z \mP)_0 (\d_Z \overline{S})_0.
\end{align}
And now formally set the radial and vertical stratification lengths
to
\begin{align}
\frac{1}{H_R} \equiv \frac{1}{\gamma}(\d_R\overline{S})_0, \qquad
\frac{1}{H_Z} \equiv \frac{1}{\gamma}(\d_Z\overline{S})_0.
\end{align}
Then the dimensional Boussinesq equations may be written as
\begin{align}
&\frac{D\u'}{Dt} = -\frac{1}{\mrho_0}\nabla P -2\Omega_0\ez\times\u - 
                 \Omega_0(q_R\,u_x +q_Z\,u_z) \ey \notag \\
&  \hskip2cm  
+\left(H_RN_R^2 \ex + H_ZN_Z^2 \ez  \right)\left(\frac{\rho'}{\mrho_0}\right), \label{boozemom}\\
&\nabla \cdot \u' =0, \\
&\frac{D}{Dt}\left(\frac{\rho'}{\mrho_0}\right) = \frac{1}{H_R} u_x' + \frac{1}{H_Z} u_z' + \Xi[\rho'].
\end{align}
Usually, we choose stratification in one direction only (either radial
or vertical), and then it is
convenient to introduce the buoyancy variable
$ \theta = H\rho'/\mrho_0$ (with units of length), 
where $H$ can be either $H_R$ or $H_Z$. As a consequence,
 the system only depends
on the buoyancy frequency and the rotation
law. 

The momentum equation in terms of total,
rather than perturbed, velocity, is
\begin{align}
&\frac{D\u}{Dt} = -\frac{1}{\mrho_0}\nabla P -2\Omega_0\ez\times\u - 
                 2\Omega_0^2(xq_R  +zq_Z ) \ex \notag \\
 & \hskip2cm +\left(H_RN_R^2 \ex + H_ZN_Z^2 \ez  
      \right)\left(\frac{\rho'}{\mrho_0}\right). \label{boozemomdim}
\end{align}
Finally, if we want the temperature perturbation then we
turn to the equation of state, which to leading order
gives
$\rho'/\mrho_0 = -T'/\overline{T}_0.$

In summary, the Boussinesq approximation requires that $\epsilon\ll 1$ and
$\mathcal{M}\ll 1$, as in the incompressible model, but it differs 
in the permitted magnitude of the fractional density perturbation. It assumes
$P'/\mP\ll\rho'/\mrho\ll 1$. 
The larger density fluctuation gives rise to
buoyancy terms in both the radial and vertical direction. Note that if
the box is placed at the midplane there is no vertical buoyancy
force. Note also that the radial buoyancy force is of order $H_Z/H_R$
smaller than the other terms and could be neglected in certain circumstances. 
The equations do not guarantee that $\rho'$ remains
small in the way defined: it is possible for the system to evolve away
from its domain of validity, just as in the incompressible box.

Finally, these equations are what Umurhan and Regev
(2004, 2008) refer to as the `small shearing box', which they derive
in an alternative manner.
 Because the
equation for their buoyancy variable arises from the continuity
equation, however, their system cannot incorporate background 
entropy gradients (hence convection) nor diabatic effects such as
cooling or thermal diffusion, and is hence far more restrictive.

\subsection{Conservation laws}

As in Section 3.1, we derive a number of conservation laws that the
flow obeys. To make life simpler, we set $q_Z=0$ and assume
stratification in only one direction, the $z$ direction. Thus
$N_R=0$. The more
general case of $x$ and $z$ stratification can be found
in Appendix B.  In addition, the fluid is taken to be adiabatic so that
$\Xi=0$. Finally, we do not treat
angular momentum, as the result is identical to that appearing in the
incompressible
case (Section 3.4.2).

\subsubsection{Energy}

By taking scalar products and manipulating, we obtain the energy result
\begin{align}
&\d_t (\tfrac{1}{2}u^2+\tfrac{1}{2}N_Z^2\theta^2)\notag \\
& \hskip1cm +\nabla\cdot\left[(\tfrac{1}{2}u^2+h+\tfrac{1}{2}\Omega_0^2x^2q_R + 
\tfrac{1}{2}N_Z^2\theta^2)\u\right]=0.
\end{align}
 The specific `thermal energy' in the
Boussinesq box is thus $\tfrac{1}{2}N_Z^2\theta^2$. 

\subsubsection{Potential vorticity}

We introduce the convenient `total' buoyancy variable
$\theta_z= H_Z\rho'/\mrho_0 + z$, which transforms the entropy
equation into 
\begin{equation}\label{eq2}
\frac{D\theta_z}{Dt}=0.
\end{equation} 
Taking the
curl of \eqref{boozemomdim} 
obtains an equation for the vorticity in the shearing
box:
\begin{equation}
\frac{D\omi}{Dt} = \omi\cdot\nabla\u - N^2_Z\nabla\theta_z
\times \ez\,, \label{eq4}
\end{equation}
where $\omi$ is the vorticity, defined in Eq.~\eqref{vort}.
This equation is saying that vorticity can be generated by
the buoyancy term, but only in a direction perpendicular to both $\ez$
(the direction of the background entropy gradient)
and to the
gradient in $\theta_z$, which suggest ways
 to construct conserved quantities.

We take the inner product of
\eqref{eq4} with $\ez$ and obtain:
\begin{equation*}
\frac{D\omega_z}{Dt} = \omi\cdot\nabla u_z,
\end{equation*}
which, with incompressibility, can then be transformed into the
conservation law:
\begin{equation}
\frac{\d \omega_z}{\d t} + \nabla\cdot\left(\omega_z\,\u - u_z\,\omi\right)=0. \label{con1}
\end{equation}
The component of the vorticity in the
direction of the stratification is always conserved.

We now consider the second direction and 
take the inner product of \eqref{eq4} with
$\nabla\theta_z$, giving
\begin{equation}
\nabla\theta_z\cdot \frac{D\omi}{Dt} 
= \nabla\theta_z\cdot\left[(\omi\cdot\nabla)\u\right]. \label{eq5}
\end{equation}
Next we take the gradient of \eqref{eq2} and find
\begin{equation}
\frac{D\nabla\theta_z}{Dt} = -(\nabla\u)\cdot\nabla\theta_z. \label{eq6}
\end{equation}
Putting these two together
obtains a second conservation law:
\begin{equation}
\frac{\d \Theta}{\d t} + \nabla\cdot(\Theta\,\u) =0, \label{con2}
\end{equation}
where the conserved quantity is defined by
$\Theta\equiv\omi\cdot\nabla\theta_z$. This we regard as the \emph{potential vorticity} in
the Boussinesq shearing box. Moreover, because of incompressibility we
have $D\Theta/Dt=0$, and $\Theta$ is materially conserved. 
Of course, Eq.~\eqref{con2} is nothing
but Ertel's
theorem in the context of Boussinesq hydrodynamics (M\"uller, 1995).

Note that a conservation law is possible even when there is some kind
of `frictional force', $\mathbf{G}$, in \eqref{boozemomdim} (see
Haynes and McIntyre 1987). This could be
viscosity or hyperviscosity. The vorticity
equation now picks up a term $\nabla\times\mathbf{G}$ on its right
hand side. Taking the inner product of the new term with $\nabla\rho$
yields
\begin{equation}
 \nabla\rho\cdot(\nabla\times\mathbf{G})
 =\nabla\cdot(\mathbf{G}\times\nabla\rho),
\end{equation}
where we have used the fact that a curl of a gradient is zero. The
modified conservation law for $\Theta$ is then
\begin{equation}
\frac{\d \Theta}{\d t} 
+ \nabla\cdot\left(\Theta\,\u + \mathbf{G}\times\nabla\rho \right)=0.
\label{con3}
\end{equation}

\section{The anelastic approximation}

The next model in our hierarchy still requires the motions to be slow
and the Mach number and thermodynamic perturbations small. However, it
extends the size of the box to cover the disc's
vertical scale height. In its strongest form it also permits $\lambda$
to reach $H_Z$.
This is the anelastic model (Ogura and Phillips 1962, Gough 1969), 
used with some success in modelling solar convection,
though less frequently in the disc context (see Barranco and Marcus
2005, 2006). Basically, an anelastic shearing box is a vertically
stratified shearing box in which the sound waves and other compressible
dynamics have been filtered out. 

As has been discussed elsewhere, it is not straightforward to formally
derive an anelastic model that satisfies the conservation laws one
would want it to. This can lead to the advent of spurious
instabilities, amongst other problems (e.g.\ Bannon 1996, Brown et al.~2012, Vasil
et al.~2013). One response has been to make ad hoc tweaks to the derived
equations in order that the conservation laws are preserved. Of
course, the relationship of the new equations to the original
 set is unclear, and even then the new systems are not conservative in
 general.
 
In this section we revisit the problems the analestic
model faces. We then show that anelastic models with additional
restrictions can be derived that possess the correct properties. 

\subsection{Classical anelastic equations}

In this section we demonstrate how to derive the `classical
anelastic equations' as presented in Gough (1969)
but in the context of an astrophysical disc (not a star). 
The model permits the characteristic lengthscale
of phenomena to reach the scale height, i.e. $\lambda\sim H_Z$ (thus
$\epsilon\sim 1$), but for simplicity we set
$H_Z\ll H_R$. The fractional perturbations in density and pressure are
assumed small, so that
\begin{equation}
\frac{\rho'}{\mrho}\sim
\frac{P'}{\mP}\sim \mathcal{M}^2 \ll 1,
\end{equation}
where we assume that the Mach number of the flow is also small.

\subsubsection{Derivation and governing equations}

The first thing to note is that the background thermodynamic
variables $\mrho$ and $\mP$ can no longer be expanded in
$z$ and then truncated at leading order. They may still, however, be
expanded and truncated in small $x$, as we assume that $\lambda \ll
H_R$. In contrast, the rotation profile $\Omega$ may be expanded in
$z$ because its characteristic scale of variation is $\sim
(H_R/H_Z)R_0 \gg \lambda$. The natural location to anchor an anelastic
shearing box is at the midplane, and thus $Z_0=0$. But then this means
$(\d_z\Omega)_0=0$ and vertical shear drops out of the problem;
the quadratic
terms in $z$ are at most $(H_Z/R_0)^2$ smaller than the leading order terms.  
To retain vertical shear, additional scaling assumptions are required
such as radial geostrophic balance (see Nelson et al.~2013).

We now rescale the variables. First, $x=H_Z x^*$, etc, and $t=
(H_Z/w)t^*$, and the perturbations follow
\begin{align}
\u'=w\u^*, \quad \rho'= \mathcal{M}^2\mrho_0 \rho^* 
 \quad 
P'=\mathcal{M}^2\mrho_0 H_Z^2\Omega^2_0 P^*,\label{anepertscl}
\end{align}
where now $\mrho_0$ should be understood as the equilibrium density 
at the midplane $Z=Z_0=0$ and at $R=R_0$.
The equilibrium quantities are written as
\begin{align}\label{goon1}
&\mrho= \mrho_0 \mrho^*(z), \quad \mP = \rho_0 H_Z^2\Omega_0^2
\mP^*(z), \\
& \d_R\mrho = \mrho_0 \frac{1}{H_R}(\d_R\mrho)^*(z), \quad
 \d_R\mP = \mrho_0 \frac{H_Z^2}{H_R}(\d_RP)^*(z),\label{goon2}
\end{align}
where it is understood that the starred quantities are the
dimensionless leading-order terms in an expansion in $x$ around $R=R_0$. They do not depend
on $x$ but they do depend on $z$, as indicated. Finally, $\ou=\lambda
\Omega_0 q_R x^* \ey$, where $q_R$ is defined in \eqref{queues}.

These scaling are substituted into the governing equations. The continuity
equation at order 1 becomes
\begin{equation}
\nabla^*\cdot(\mrho^* \u^*) =0.
\end{equation}
The other terms are either $\mathcal{M}^2$ or $H_Z/H_R$ smaller. 
The momentum equation to leading order is
\begin{align}
&\frac{D\u^*}{Dt^*}=
 -\frac{1}{\mrho^*}\nabla^*P^*
 +\frac{\d^*_z\mP^*}{\mrho^*}\left(\frac{\rho^*}{\mrho^*}\right)\ez \notag\\
& \hskip2cm -2\Ro^{-1}\ez\times\u^* - \Ro^{-1}q_R\,u_x^* \ey,
\end{align}
where Ro$=w/(\lambda\Omega_0)$ is the Rossby number.
Again we have dropped a term a factor $H_Z/H_R$ smaller than the
others (the radial gradient of the background pressure).
Also,
$(\d^*_z\mP^*/\mrho^*)$ may be replaced by $z^*$, on using
the vertical hydrostatic balance equation. 

The entropy equation can be tackled by first linearising $S$
in $P^*$ and $\rho^*$, which sets the entropy perturbation as
\begin{equation} \label{anelS}
S^*= \frac{P^*}{\mP^*} -\gamma\frac{\rho^*}{\mrho^*}.
\end{equation}
Likewise, the heating and cooling terms can
be expanded. The resulting entropy equation is written in terms
of density and pressure or, more compactly, as
\begin{align}
\frac{DS^*}{Dt^*} = - u_z^*\d_z^*\overline{S}^* +(\d_\rho\Xi)^*\rho^* +
(\d_P\Xi)^* P^*,
\end{align}
where we have packaged heating and cooling into the single function
$\Xi(\rho,\,P)$.
Returning to total velocity variables and putting things in
dimensional form, we get the set:
\begin{align}
&\nabla\cdot(\mrho\u)=0, \label{aniw1}\\
&\frac{D\u}{Dt}=
 -\frac{1}{\mrho}\nabla^*P'
 -z\Omega_0^2\left(\frac{\rho'}{\mrho}\right)\ez \notag\\
& \hskip2cm -2\Omega_0\ez\times\u - \nabla\Phi_T,\label{aniw2}\\
& \frac{DS'}{Dt} = -u_z \d_z\overline{S} +
(\d_\rho\Xi)\rho' +
(\d_P\Xi) P', \label{aniw3}\\
& S'= \frac{P'}{\mP} -\gamma\frac{\rho'}{\mrho}.\label{aniw4}
\end{align}
where $\Phi_T= \Omega_0^2q_R x^2$.

\subsubsection{Conservation issues}

As has been pointed our for some time, 
the anelastic approximation struggles with
conserving energy. See Brown et al.~(2012) for some discussion
of this issue. Certainly, as they stand, equations
\eqref{aniw1}-\eqref{aniw4} do not conserve total energy. 

First by taking the scalar product of \eqref{aniw2} with $\u$, one obtains
an equation for the kinetic energy. With a bit of manipulation this is
\begin{align}
&\d_t(\tfrac{1}{2}\mrho u^2)
+\nabla\cdot\left[\u\left(\tfrac{1}{2}\mrho u^2+\mrho\Phi_T + P'
  \right)
 \right]\notag \\ 
 &\hskip2cm = -P'\nabla\cdot\u +\frac{\rho'}{\mrho}\u\cdot\nabla\mP\notag \\ 
 &\hskip2cm =-P' {\mP}^{-1/\gamma}\nabla\cdot( {\mP}^{1/\gamma}{\bf u}) -\frac{S'}{\gamma}\u\cdot\nabla\mP.
\label{aniec} \end{align}
For demonstration purposes, we assume that
 the background state is barotropic so that ${\overline P} = {\overline P}(\overline S)$ and thus
$\nabla {\overline P}= (d{\overline P}/d{\overline S}) \nabla {\overline S}.$
Using this and (\ref{aniw4}) we obtain the total energy equation
\begin{align}
&\d_t(\tfrac{1}{2}\mrho u^2+{\cal V})
+\nabla\cdot\left[\u\left(\tfrac{1}{2}\mrho u^2+\mrho\Phi_T +{\cal V}+ P'
  \right)
 \right] \notag \\ 
  & =\frac{-P'}{ {\mP}^{1/\gamma}}\nabla\cdot( {\mP}^{1/\gamma}{\bf u})-\frac{S'}{\gamma} \frac{ d{\mP}}{d{\overline S}} 
\left[(\d_\rho\Xi)\rho' +
(\d_P\Xi) P'\right]\notag \\ 
&\hskip1cm -\frac{\mrho S'^2}{2\gamma}\frac{d }{d{\overline S}}
\left(  \frac{1}{\mrho}   \frac{ d{\mP}}{d{\overline S}}\right)
\u\cdot\nabla{\overline  S}.
\label{aniec2}
 \end{align}
Here ${\cal V}\equiv- S'^2(d{\mP}/d{\overline S})/(2\gamma)$
plays the role of thermal energy.  
We remark that for linear adiabatic perturbations 
$S' = -\xi_zd{\overline S}/{dz},$  where $\xi_z$ is the 
vertical component of the Lagrangian displacement, 
and accordingly
\begin{align}
&{\cal V}= \frac{1}{2}\mrho\xi_z^2N_Z^2,
 \end{align}
which being a quadratic form in the perturbation,
 represents the potential energy associated with linear buoyant motions.

Each of the three terms on the right hand side of (\ref{aniec2})
destroys conservation.
The first term is the most serious as it is non-zero when the
background entropy gradient is non-zero, and thus only vanishes for
homentropic equilibria. The term is associated with a failure to conserve energy properly
even for adiabatic linear waves, and thus
wave-action conservation
for those waves is incorrectly represented. The second term vanishes
for adiabatic motions,
and is hence unproblematic,
while the third term is third order in the amplitude of the perturbations.
This third term does not affect energy conservation for linear perturbations
but could cause departures on a time scale that scales inversely
with the amplitude of those perturbations.

Similar problems afflict most 
formulations of the anelastic  equations.
For example, the anelastic scheme introduced by Bannon (1996) and
employed by Barranco \& Marcus (2005, 2006), replaces the
entropy perturbation Eq.~\eqref{anelS} with an ad hoc prescription
that is only strictly true for a vertically homentropic background.
Furthermore, they show even then that energy conservation only holds in the
special case of an \emph{isothermal} background. We are aware of no
anelastic system that is conservative in general. If indeed conservative
anelastic models are confined to isothermality, their utility is
greatly reduced; they can no longer reliably describe
convection, nor the influence of stratification on wave propagation,
dynamo action, etc. 

In the next subsection, we derive a set of conservative anelastic equations
that retains diabaticity. The price to be paid, however, is a
restriction on
the characteristic wavelength of the phenomena, which must be significantly less
than $H_Z$ (as in the Boussinesq approximation), yet we permit the
flow to range over a domain of size $H_Z$. Physical problems well
suited to this model include: wave packets propagating upward from the
disc midplane to the surface, where they may suffer refraction,
wave channelling, or breaking (Bate et al.~2002), and small-scale disc turbulence
localised to certain altitudes, such as MRI in the disc surface
(Fleming and Stone 2003), or
convection in a limited range of convectively unstable layers (Stone
and Balbus 1996). The system derived below bears a close
similarity to the `pseudo-incompressible' equations, first presented
by Durran (1989) in the context of atmospheric dynamics. We, however, must
enforce the additional lengthscale restriction because of the strong
differential rotation exhibited by astrophysical disks (absent in most
atmospheric and planetary applications). This dynamical
feature is uniquely dominant in disks, and presents a considerable obstacle
to the derivation of consistent anelastic models in that context.

\subsection{A conservative anelastic model}

We assume that the phenomena are spread across a domain of spatial
extent $\sim H_Z,$ meaning that 
$ -\mathcal{O}(H_Z )< (x,y,z) < \mathcal{O}( H_Z)$. However, we
impose the condition that the characteristic radial and vertical 
length scales of the
phenomena  under study are much smaller, so that $\lambda\ll H_Z$, or in
other words $\epsilon\ll 1$.
 The scale of these motions in the direction of shear, $y,$ on the
 other hand,
 we let equal the vertical scale height
which makes the motions almost axisymmetric (the tight-winding
approximation). 
It is necessary to enforce short scales, radially at
the very least, because across $\sim H_Z$ the background shear
velocity is supersonic, and large-scale disturbances rapidly wind up
into smaller scale structures. A related issue is the scaling of
the background pressure, Eq.~\eqref{pressure}, which differs from that
appearing in planetary and stellar contexts. 

In contrast to the classical anelastic approximation, we scale the
fractional thermodynamic perturbations according to
\begin{equation}
 \frac{\rho'}{\mrho}\sim \epsilon, \qquad \frac{P'}{\mP}\sim
\mathcal{M}^2.
\end{equation}
With this scaling $P'/\mP\sim\epsilon^2 \Ro^2$,
so that for Rossby numbers of order unity we are
 assured $ P'/\overline{P}\ll\rho'/\overline{\rho}.$

We introduce dimensionless independent variables,
$x=\lambda x^*,$   $ z=\lambda z^*,$ and $t=
(\lambda/w)t^*$, 
and scale the dependent variables through
\begin{align}
\u'=w\u^*, \quad \rho'= \epsilon\mrho_0 \rho^*, \quad
P'=\mathcal{M}^2\mrho_0 H_Z^2\Omega^2P^*,
\end{align}
where now $\mrho_0$ should be understood as the equilibrium density 
at the midplane and at $R=R_0$.
The equilibrium quantities are written as earlier in Eqs \eqref{goon1}
and \eqref{goon2}.

\subsubsection{The continuity equation}
These scaling are inserted into the
continuity
equation which  becomes, without approximation,
\begin{equation}
\epsilon \left(\frac{\partial \rho^*}{\partial t^*}
 +\frac{q_R x}{\Ro} \frac{\partial \rho^*}{\partial y} 
+\nabla^*\cdot(\rho^* \u^*) \right)  
+   \nabla^*\cdot(\mrho^* \u^*) =0.
\label{anelcont}\end{equation}
 Here $\nabla^*\equiv {\bf e}_x\partial/\partial x^* + {\bf
   e}_z\partial/\partial z^* 
+{\bf e}_y\lambda \partial/\partial y.$  
We have been careful to retain dimensional $x$  where it appears in equation~\eqref{anelcont}
 because the radial size of the domain may greatly exceed
$\lambda$. In addition, the $y$ derivative in the del operator is kept
dimensional for the moment.
 Note that the term proportional to $\epsilon$ in Eq.~\eqref{anelcont} 
is small and one might 
expect to be able to neglect it. This is appropriate if disturbances with scales $\sim \lambda$ are considered
in a domain of comparable size.
However, if  we wish to consider disturbances propagating over scales $H_Z \gg \lambda,$
doing so may be invalid and this term is retained for now.
The contribution ${\bf e}_y\lambda \partial/\partial y$ in $\nabla^*$  can be argued to be small for disturbances
of scale $H_Z$ in the $y$ direction. However, on account of the second term in (\ref{anelcont}), the dependence on $y$ may not be neglected except for axisymmetric disturbances 
so we shall also retain this term.  

By making use of the identity
\begin{equation}
\nabla^*\cdot(\rho^* \u^*)=   \frac{\rho^*}{\mrho^*} \nabla^*\cdot(\mrho^* \u^*) +\mrho^*\u^*\cdot\nabla^* \left(\frac{\rho^*}{\mrho^*}\right)
\end{equation}
equation
(\ref{anelcont})  may be rewritten in the form
\begin{equation}
\frac{\lambda \mrho^2}{\rho\mrho_0 H_Z} \left(\frac{\partial }{\partial t^*} +\frac{q_R x}{\text{Ro}} \frac{\partial }{\partial y} +\u^*\cdot\nabla^* \right)\left(\frac{\rho^*}{\mrho^*}\right)  +   \nabla^*\cdot(\mrho^* \u^*) =0,
\label{anelcont1}
\end{equation}
where we recall that $\rho = (\epsilon\rho^*+ \mrho^*) \mrho_0.$

\subsubsection{The entropy equation}
We begin by recognising that, for a simple ideal gas, 
the density, pressure and  entropy are related by
\begin{equation}
\rho=P^{1/\gamma}\exp(-S/\gamma).
\end{equation}
  Writing $S ={\overline S} +S',$
where
${\overline S}$  is the equilibrium entropy and $S'$ is the perturbation, we have
\begin{equation}
1+\frac{\rho'}{\mrho} =\left(\frac{\mP+P'}{\mP}\right)^{1/\gamma}\exp(-S'/\gamma).\label{anent}
\end{equation}
In our ordering scheme the pressure perturbation is of 
higher order than the relative density perturbation. For this  reason it will be neglected.
The removal of pressure fluctuations in this way filters out sound waves 
and thus leads us to an anelastic approximation.
 Adopting dimensionless variables, the entropy perturbation is then related to the density perturbation through
\begin{equation} 
\exp(-S^*/\gamma) = 
 1+\epsilon\frac{\rho^*}{\mrho^*},\label{entr}
\end{equation}
where we have written  $S^*  \equiv S'.$

 Neglecting the pressure perturbation therein,  the heating/cooling term is
 expressed in terms of $\rho^*/\mrho^*$ only. 
The  entropy equation will then  only contain
the  density perturbation,  taking the form
\begin{align}
 \left (\frac{\partial  }{\partial t^*} +\frac{q_R x}{\text{Ro}} \frac{\partial }{\partial y} +\u^*\cdot\nabla^* \right) S^*
 = 
 -u_z^*\d_z^*\overline{S} 
 +(\d_\rho\Xi)^*\rho^*. \label{analentr}
\end{align}
We emphasise that the only approximations made up to now are the neglect of the pressure perturbation
together with the linearization of the heating/cooling term . 
We go on to use (\ref{entr}) to eliminate $\rho^*$  in equation  (\ref{anelcont1}); the latter then takes the form
\begin{equation}
\frac{ \mrho}{\gamma \mrho_0} \left(\frac{\partial }{\partial t^*} +\frac{q_R x}{\text{Ro}} \frac{\partial }{\partial y} 
+\u^*\cdot\nabla^* \right)S^*  =   \nabla^*\cdot(\mrho^* \u^*).
\label{anelcont2}
\end{equation}
We remark that, on account of the smallness of the neglected pressure perturbation on the left hand side,
 the above equation incorporates all corrections of order $\epsilon$ to the
dominant term on the right  hand side.

\subsubsection{The momentum equation}
The only approximation made in the momentum equation is the
neglect of the pressure perturbation except where it appears as a gradient.
We shall work directly with the unscaled nonlinear equation, which takes the form
\begin{align}
&\rho \frac{D\u}{Dt}=
 -\nabla P
 -2\rho\Omega_0\ez\times\u -\rho \nabla\Phi,\label{ani20}
\end{align}
where the potential  $\Phi =\Phi_Z +\Phi_T,$ with  $\Phi_Z=
\Omega_0^2z^2/2$ and  
 $\Phi_T= {q_R}\Omega_0^2x^2$.
We set $P= {\overline P} +P'$ with $ d{\overline P}/dz =\mrho d\Phi_Z/dz$ which
is just the condition of vertical hydrostatic equilibrium. 
Equation (\ref{ani20}) then  becomes
 \begin{align}
&\rho \frac{D\u}{Dt}=
 -\nabla P'
 -2\rho\Omega_0\ez\times\u -\rho \nabla\Phi_T    -(\rho-\mrho) \nabla\Phi_Z,\label{ani21}
\end{align}
Anticipating  what is needed to obtain a system that conserves energy,
we \emph{modify} equation (\ref{ani21}) so that it  becomes
 \begin{align}
&\rho \frac{D\u}{Dt}=
  -{\overline P}^{1/\gamma}\nabla\left( \frac{P'}{{\overline P}^{1/\gamma}}\right)\nonumber \\
&\hspace{1cm}  -2\rho\Omega_0\ez\times\u -\rho \nabla\Phi_T    -(\rho-\mrho) \nabla\Phi_Z.\label{ani22}
\end{align}
Note that  (\ref{ani21}) and (\ref{ani22}) differ by  a term on the RHS that is $\propto\,  P'$ 
and is of order $\epsilon$ smaller than the term involving the
gradient of $P'$. Naively, the two equations may be considered identical, to leading order. 
An objection one may raise, however, is that this procedure permits
the addition of arbitrary terms of the same order and we end up with a
set of equations lacking uniqueness. This is not a problem here. 
The term added to the right hand side is
 \begin{equation}
 \frac{P'}{\gamma \mP}\nabla\mP  =  -\frac{P'\mrho}{\gamma \mP}\nabla\Phi_Z,
 \end{equation}
 which can be viewed as  supplying an additional contribution to  $\rho - \mrho.$
 This contribution turns out to be  $\propto\, P'$ and was in fact
 dropped when approximating equation (84) by (85).
The procedure hence reincorporates this contribution correct to linear order in the density and pressure perturbations.
 This justifies the modification.

 \subsubsection{Final anelastic governing equations}
 Putting the equations in 
dimensional form and eliminating $S'$ from (\ref{analentr}) and
(\ref{anelcont2}),
 obtains the set:
\begin{align}
& \nabla\cdot(\mrho\u) = -\frac{{\overline \rho}}{\gamma}\left( u_z \d_z\overline{S} -
(\d_\rho\Xi)(\rho-\mrho) \right)
, \label{ani3}\\
&\rho \frac{D\u}{Dt}=
  -{\overline P}^{1/\gamma}\nabla\left( \frac{P'}{{\overline P}^{1/\gamma}}\right)\nonumber \\
&\hspace{1cm}  -2\rho\Omega_0\ez\times\u -\rho \nabla\Phi_T    -(\rho-\mrho) \nabla\Phi_Z.\label{anise}\\
&\frac{D\rho}{Dt}  = -\rho\nabla\cdot \u.
 \label{ani4}
\end{align}
These yield five  equations for the three velocity components,  $\rho$ and $P'.$
The pressure perturbation, $P',$ is determined after applying (\ref{ani3}) as a constraint 
condition on the velocities in the same way as would be done in an incompressible 
model. In this case the constraint (\ref{ani3}) replaces the condition $\nabla\cdot \u=0$
in the incompressible case, and $\nabla\cdot(\mrho\u)=0$ in the
classical anelastic model.

 An important aspect of the new equations  is that they
  yield conservation laws for entropy,
potential 
vorticity and total energy when heating  and/or cooling is absent.
In this case, the constraint equation (\ref{ani3}) becomes
\begin{equation}
\nabla\cdot({\overline P}^{1/\gamma}\u)=0, \label{aneincomp}
\end{equation}
which is in a form familiar from the pseudo-incompressible
approximation (Durran 1989, Vasil et al.~2013).

\subsubsection{Conservation of energy}
Taking the scalar product of (\ref{ani22}) with ${\bf u},$
while making use of (\ref{aneincomp}) and vertical hydrostatic equilibrium,
gives us the conservation  law for the   energy of the system,  in the form
 \begin{align}
&\d_t\left(\rho\left(\tfrac{1}{2}u^2 +  \Phi_T +\Phi_Z + \frac{ \mP}{ \rho(\gamma-1)}\right)\right)+
\notag\\
&\nabla\cdot\left[\u\left(\tfrac{1}{2}\rho{u}^2+\rho(\Phi_T +\Phi_Z)+ P'
 +\frac{\gamma\mP}{\gamma-1} \right)\right] =0.
\label{anelecon}
 \end{align}
The quantity  $\mP/[(\gamma -1)\rho]$  on the left hand plays the role of the internal energy per unit mass.
It involves $\mP$ rather than the usual  $P$ on account of the neglect of $P'$ in the entropy equation.
Owing to the cancellation of $\rho$ it ultimately makes no contribution. However, we have  included 
it in order to relate to the  general case.

\subsubsection{Conservation of entropy}
Combining (\ref{ani3}) with (\ref{aneincomp}) we obtain
\begin{equation}
\frac{D}{Dt}\left(\frac {\rho}{{\overline P}^{1/\gamma}}\right)=0. \label{anelentc}
\end{equation}
This is a statement of the conservation of entropy.
But note that  
${\overline P}^{1/\gamma}$ occurs rather than the expected
$ P^{1/\gamma}. $ This is because the Eulerian pressure  
perturbation is assumed to be negligible in the anelastic model.

\subsubsection{Conservation of potential vorticity}
Dividing  (\ref{ani22}) by $\rho,$ taking the curl and making use of the continuity equation
(\ref{ani4})   we obtain
\begin{align}
&\rho \frac {D}{Dt}\left( \frac{\omi}{\rho}\right) 
-\omi\cdot\nabla\u   = \nonumber\\
&\nabla\left({\overline P}^{1/\gamma}/\rho \right)
\times\left(  \frac{\mrho} {{\overline P}^{1/\gamma}}\nabla\Phi_Z       - \nabla( P'/{\overline P}^{1/\gamma} )  \right),
\end{align}
where $\omi= \nabla\times\u + 2\Omega_0\ez $ is the
absolute vorticity. 
Making use of (\ref{anelentc}) we obtain the conservation of  potential vorticity
\begin{align}
&\frac {D}{Dt}\left[\left( \frac{\omi}{\rho}\right)\cdot \nabla\left(\frac {\rho}{{\overline P}^{1/\gamma}}\right)\right]  =0 
\end{align}
Note that this conservation law survives the introduction of viscous
forces as indicated in Section 4.4.2.

\section{Compressible dynamics}

Finally we deal with local models that fully incorporate compressible
motions, and thus describe fast phenomena such as sound waves and
transonic turbulence. In the previous `slow' approximations we set the
Mach number to be small, but now suppose
 $\mathcal{M}$ can take any value. In addition, we do
not initially specify the length scales over which the dynamics will
manifest, and thus $\epsilon = \lambda/H_Z$ 
will be free for the time being. 
The thermodynamic 
perturbations are assumed to  be such that
\begin{align}
\frac{\rho'}{\mrho}\sim 1, \qquad \frac{P'}{\mP}\sim 1.
\end{align}
and now the equations are set up to capture large fluctuations in density and
pressure.

Enforcing radial locality, we expand the background in small
$x$ and retain the leading order terms. 
The equilibrium, however, retains its full dependence on $z$ at first. 
We next scale
the background fields according to Eqs \eqref{youse} and \eqref{goon1}-\eqref{goon2}.
Perturbation scalings are given by (\ref{anepertscl})
 but with ${\cal M}$ set equal to unity.
 Finally, $x=\lambda x^*$, etc, and $t=(\lambda/w)t^*$ where
a star indicates a dimensionless order 1 quantity. 

We insert the scaled form of the dependent and independent variables
into the governing equations \eqref{nl1}-\eqref{nl3}. The 
continuity equation becomes
\begin{equation} \label{fullcont}
\frac{D\rho^*}{Dt^*} = -(\mrho^*+\rho^*)\nabla^*\cdot\u^*
-\epsilon u_z^*\d_z^*\mrho^* -\epsilon\frac{H_Z}{H_R}u_x^*(\d_R\,\mrho)^*.
\end{equation}
Here we have made use of  (\ref{eqmscale1}) and (\ref{eqmscale2}).
Similarly for the momentum
equation we have
\begin{align}
&\frac{D\u^*}{Dt^*} =
-\frac{\mathcal{M}^{-2}}{\mrho^*+\rho^*}\nabla^*P^*
-2\Ro^{-1}\ez\times\u^* \notag \\
&  \hskip2.5cm- \Ro^{-1}(q_R\,u_x^*+q_Z\,u_z^*)\ey \notag\\
& \hskip0.5cm        +
        \epsilon\mathcal{M}^{-2}\frac{\rho^*}{\mrho^*(\mrho^*+\rho^*)}
        \left(\d_z^*\mP^*\ez
         +\frac{H_Z}{H_R}\d_R^*\mP^*\ex \right), \label{fullmom}
\end{align}
where again use has been made of (\ref{eqmscale1})
 and (\ref{eqmscale2}). In the above, Ro$=w/(\lambda\Omega_0)$ is the
 Rossby number.
We next analyse these equations in the two relevant limits for
compressible flow.

\subsection{The `small' compressible shearing box}

The next assumption we make is that the compressible phenomena of
interest
takes place on small scales, so that $\lambda\ll H_Z$,
i.e. $\epsilon\ll 1$. We may then expand the background
thermodynamic variables
in small $z$, and to leading order they become constant. 
The continuity equation is, to leading order in $\epsilon$,
\begin{equation}
\frac{D\rho^*}{Dt^*} = -(\mrho^*+\rho^*)\nabla^*\cdot\u^*,
\end{equation}
while the momentum equation is obtained by making use of
(\ref{i2}), ({\ref{eqvel}) and (\ref{ueq}):
\begin{align}\label{momstandard}
&\frac{D\u^*}{Dt^*} =
-\frac{\mathcal{M}^{-2}}{\mrho^*+\rho^*}\nabla^*P^*
-2\Ro^{-1}\ez\times\u^* \notag \\
&  \hskip2.5cm- \Ro^{-1}(q_R\,u_x^*+q_Z\,u_z^*)\ey. 
\end{align}
We have not explicitly constrained the velocity scale $w$ yet. If we
insist that it is of order the sound speed then $\mathcal{M}\sim 1$,
and $w\sim H_Z\Omega_0$, but this immediately means that 
$\Ro^{-1}\sim \epsilon$ and hence the shear and rotation terms drop
out! Small-scale fast disturbances are unaware they exist in a
shearing box. As the compressible model may also describe slower
phenomena,
in addition to transonic flow,
and the only approximation made
so far is that $\epsilon \ll 1,$ we keep all the terms in 
\eqref{momstandard} for the moment.
Finally, the entropy equation is simply
$DS^*/Dt^* = \Xi^*,$
where heating and cooling have been incorporated into the single
function $\Xi$,
and the background gradients in the entropy are subdominant and
do not appear. 
 
\subsubsection{Final compressible equations}

In dimensional form and using total variables rather than
perturbations, we have the set
\begin{align}
&\frac{D\rho}{Dt} = -\rho\nabla\cdot\u, \label{comp1}\\
&\frac{D\u}{Dt} = -\frac{1}{\rho}\nabla P -2\Omega_0\ez\times\u
-\nabla\Phi-(2\Omega_0^2q_Z)z \ex, \label{comp2}\\
& \frac{DS}{Dt} = \Xi, \label{comp3}
\end{align}
where the tidal potential is $\Phi= \Omega_0^2q_R x^2$.
We note that to maintain consistency in the ordering scheme,
the background density and pressure are taken to be constants
and there is no vertical stratification.
This is because terms involving the gradient of the background state variables 
have vanished as a result of the assumption $\epsilon \ll 1.$

A few comments about this `small box' compressible model.
First, especially in numerical simulations, these equations are frequently
employed to describe phenomena on  length scales of order the vertical scale height
$H_Z$, which is strictly outside its range of validity. 
Indeed, many of the seminal MRI simulations were
undertaken in computational domains equal to $H_Z$ or larger (Hawley
et al.~1995, etc). 
 If we are tracking phenomena with length scales $\sim H_Z,$ then we
should avoid using this approximation. 
However, the failure to do so is harmless for many applications, 
and the results obtained still instructive. 

Second, in a number of simulations, especially those  studying the
subcritical baroclinic instability,
background gradients in pressure are included in the momentum equation
(the last terms in \eqref{fullmom}), despite $\epsilon\ll 1$. Doing so turns out to be far from
harmless, however, as they give rise to
spurious overstabilities. This is discussed further in Section 6.4.

\subsection{The `large' (vertically stratified) shearing box}

We next allow $\lambda$  to be of order $H_Z$ and accordingly set $\epsilon=1$.
In addition, the velocity scale is set to be $\Omega_0H_Z$  so that ${\cal M}=1.$
 As in the anelastic model earlier,  
 our shearing box sits at the midplane, so that
$Z_0=0$ and hence $q_z=0$. When we expand the background flow, we obtain
$\ou = \Omega_{0}q_R x \ey $ to leading order. Vertical shear is
difficult to justify, as the next order term (quadratic in $z$) in
$\ou$ is $(H_Z/R_0)^2$ smaller than the other terms (and even smaller than
the omitted cylindrical terms). 
Unsurprisingly, Lin and Youdin (2015)
report the emergence of spurious instabilities when it is included
that may be traced back
to this inconsistency.

 Because $z$ ranges over $H_Z$, we retain the explicit
variation of $\mrho$ and $\mP$ with $z$, but Taylor expand them to leading
order in $x$.  In order to perform the latter operation we
must assume that $H_R \gg H_Z$ as $x\sim H_Z$. As a consequence, the equilibrium 
radial-gradient terms
in \eqref{fullcont} and \eqref{fullmom} must be dropped to maintain
consistency: unlike the Boussinesq approximation, there is no
flexibility when it comes to
incorporating \emph{constant} background radial gradients. If radial gradients
are sufficiently sharp, their full $x$ structure must be retained, as
is done in local models of slender tori, narrow rings, and localised
density bumps  (not treated
in this paper; see Goldreich et al.~1986 and
Narayan et al.~1987).

The continuity equation \eqref{fullcont} is then
\begin{align}
\frac{D\rho^*}{Dt^*} = -\left[\mrho^*(z)+\rho^*\right]\nabla^*\cdot\u^*
- u_z^*\d_z^*\mrho^*,\label{fullconty}
\end{align}
where we have dropped terms of order $H_Z/H_R$. 
The momentum equation becomes
\begin{align}
&\frac{D\u^*}{Dt^*} =
-\frac{1}{\mrho^*+\rho^*}\nabla^*P^*
-2\Ro^{-1}\ez\times\u^* \notag \\
&  \hskip1cm- \Ro^{-1}q_R\,u_x^*\,\ey +
       \frac{\rho^*}{\mrho^*(\mrho^*+\rho^*)}
        \d_z^*\mP^*\ez, \label{fullmommy}
\end{align}
where again we have dropped the radial gradient in the background
pressure as subdominant. Finally, these equations in dimensional form
and in total variables are
identical to Eqs \eqref{comp1}-\eqref{comp3}, but with $q_z=0$, the tidal
potential
$ 
\Phi=\Omega^2_0q_R
x^2+\frac{1}{2}\Omega_0^2 z^2,$  and
the background no longer uniform
but dependant on $z.$ We point out that this model corresponds to what
Umurhan and Regev (2008) call the `large shearing box'.

\subsection{Conservation laws}

In both compressible shearing boxes the conservation laws of energy and potential vorticity
for inviscid adiabatic flows
are relatively familiar (see  e.g.\ Ogilvie 2016). 
Introducing the specific internal energy $e=P/(\gamma-1)$,
the former may be expressed as
\begin{equation}
\d_t(\tfrac{1}{2}\rho u^2 + \rho\,e)
+\nabla\cdot\left[\rho\u\left(\tfrac{1}{2}u^2 + \Phi+e)+P\u\right)\right]=0.
\end{equation}
where the potential $\Phi$ is either $\Omega_0^2q_R
x^2$ or  $\Omega_0^2q_R
x^2 +\frac{1}{2}\Omega_0^2 z^2$. 

If $\mathcal{A}$ is any materially conserved quantity
that is a function of the thermodynamic variables
(such as the  entropy), i.e. it satisfies 
$D\mathcal{A}/Dt=0$, then it follows that the quantity
$\Theta=(\omi\cdot\nabla \mathcal{A})/\rho$ also satisfies
$D\Theta/Dt=0$, and is thus a generalisation of the potential
vorticity.

\subsection{Spurious instabilities}

In the astrophysical literature one can find examples of
small compressible shearing boxes
that include a constant gradient in the thermodynamic background; this
is in order
to either drive radial or vertical convection or in a misguided
attempt at completeness. Such terms are formally
subdominant, but when included these terms are dangerous, as
we now explore.

Consider a small box model of an isothermal gas with perturbation equations
\begin{align}
&\frac{D\rho'}{Dt} = -(\mrho_0+\rho')\nabla\cdot\u', \\
&\frac{D\u'}{Dt}  = -\frac{\nabla P'}{\mrho_0+\rho'}
-2\Omega_0\ez\times\u' -\Omega_0q_R u_x'\ey  \notag\\
& \hskip3cm +\frac{\rho'}{\mrho_0(\mrho_0+\rho')}(\d_Z\mP)_0\ez.\label{si}
\end{align}
Here $\mrho_0$ and $(\d_z\mP)_0$ are constants, and $\rho=c^2P$, with
$c$ the isothermal sound speed. 
These are essentially Eqs  (\ref{fullconty}) and (\ref{fullmommy}) but with
an extra constant term (formally subdominant)
 involving a background vertical pressure gradient so as
to allow for possible buoyancy effects.

If we linearise these equations and
seek normal modes of the type $\text{e}^{\text{i}k_z z-\text{i}\omega
  t}$,
where $\omega$ is a possibly complex frequency and $k_z$ is a real wavenumber,
then the vertical and horizontal oscillations decouple. The former
possess the dispersion relation
\begin{equation}\label{spurious1}
\omega^2 = k_z^2c^2 +\text{i}k_z\frac{(\d_Z\mP)_0}{\mrho_0}
\end{equation}
and result in growing sound waves, on account of the second
imaginary term on the right side. This term arises explicitly from
inclusion of the constant (and subdominant) vertical gradient in
pressure. Note that the instability occurs for \emph{all}  vertical   wave numbers.

Let us next include a constant \emph{radial} gradient,  but no  vertical
gradient, and  replace  the last term in \eqref{si} with 
$\rho'/[\mrho_0(\mrho_0+\rho')](\d_R\mP)_0\ex$. Again we linearise
but now examine modes $\propto \text{e}^{\text{i} k_x x
  -\text{i}\omega t}$, where $k_x$ is a real radial wavenumber. 
The ensuing dispersion relation is
\begin{equation} \label{densityunst}
\omega^2 = \kappa^2 + \text{i}k_x\frac{(\d_R\mP)_0}{\mrho_0} + k_x^2c^2,
\end{equation}
where the epicyclic frequency is defined as
$\kappa^2=2\Omega_0^2(2+q_R)$. Equation \eqref{densityunst} describes
classical density waves, but these are growing because of the
imaginary second term, which issues from the 
(subdominant) background radial
gradient. As above, the instability attacks \emph{all} radial scales.

It is straightforward to show these two instabilities
are spurious by simply examining the H\o iland criteria (e.g. Ogilvie
2016). 
We think it useful instead to tackle the two linear
problems directly in
semi-global geometries, thereby demonstrating instabilites fail to
appear when
background gradients are incorporated in full. 

To deal with the first example we turn to
the large compressible shearing
box, which exhibits the disk's full vertical structure.
In this case, the
background equilibrium is 
$\mrho=\mrho_0\text{exp}\left[-z^2/(2H^2)\right],$
where $H=c/\Omega_0$. We perturb this with modes 
$\propto \text{e}^{-\text{i}\omega t}$ and depending only on $z$.
The problem reduces to a single equation for the
perturbed enthalpy, $h=c^2\rho'/\mrho$:
\begin{equation}
\frac{d^2h}{dz^2} -\frac{z}{H^2}\frac{dh}{dz} + \frac{\omega^2}{c^2}h =0.
\end{equation}
This is the Hermite equation. For sensible solutions at $z=\pm\infty$,
the dispersion relation is
$\omega^2 = n\Omega_0^2,$
where $n$ is a positive integer. This equation should be compared
with Eq.~\eqref{spurious1}. With the identification $k_zH=\sqrt{n}$,
the first terms in each expression agree, but not the second: 
when the problem is done without
approximation there is no instability.

It is a little more involved to show that the radial instability is
spurious. To do so
we investigate the modes of a global cylindrical disc with background
structure $\mrho=\mrho(R)$, $\mP=\mP(R)$. We take the modes to be
isothermal, to depend only on $R$, and to be $\propto
\text{e}^{-\text{i}\omega t}$. The governing linearised equations are
\begin{align}
&-\text{i}\omega u_R' = -\frac{c^2}{\mrho}\d_R\rho'-2\Omega u_\phi' +
\frac{\d_R \mP}{\mrho^2}\rho', \\
&-\text{i}\omega u_\phi' = \frac{\kappa^2}{2\Omega}u_R', 
\qquad -\text{i}\omega \rho' = -\frac{1}{R}\d_R(R\mrho u_R'),
\end{align}
where the epicyclic frequency in the global problem is $\kappa^2=(2\Omega/R)d (R^2 \Omega)/dR$.
These equations may be manipulated into a single ODE for the dependent
variable $W=R\mrho u_R'$,
\begin{equation}\label{WW}
\mrho\frac{d}{dR}\left[\frac{c^2}{\mrho R}\frac{dW}{dR} \right]
 + \frac{1}{R}(\omega^2-\kappa^2)W=0,
\end{equation}
and assume the innocuous condition that $W=0$ at
the radial boundaries of the disc. 
Multiplying \eqref{WW} by the complex conjugate of
$W/\mrho$ and integrating over radius,
one can obtain
\begin{equation} \label{variational}
\omega^2 = \frac{\int  \kappa^2\, f|W|^2 dR}{\int f\,|W|^2 dR} 
  + c^2 \frac{\int f|dW/dR|^2 dR}{\int f\,
  |W|^2 dR},
\end{equation}
where $f= 1/(\mrho R).$ 
Because $f$ is always positive, both terms in
\eqref{variational} must also be positive. They, in fact, correspond
to the first and third terms in Eq.~\eqref{densityunst};
there is no equivalent
to the imaginary term. It follows that
$\omega$ is real and no instability occurs, in contradiction to 
Eq.~\eqref{densityunst}.

These demonstrations show clearly that when the problem is done
correctly in a global or quasi-global setting (in which the background
equilibrium is accounted for without approximation) the local instabilities vanish.
They are spurious. The conclusion is that it is not always safe
to include subdominant terms: they may not always offer small
corrections to the outcome but something radically incorrect that on
intermediate to long times may overwhelm a simulation.  

\subsection{Wave action  and energy conservation for small perturbations} 

The problems discussed above issue from
the fact that the fundamental conservation laws for wave action and energy,
applicable to small perturbations, break down
when spatial variations in the background state are incorporated incorrectly.
In fact, these conservation laws exclude instabilities of the
kind presented in the previous subsection.
 
In the absence of heating and cooling, 
conservation laws issue from the fact that 
the equations of motion can be derived from a stationary
action principle.
The principle is
applied to the action
\begin{equation}
\mathcal{S}[{\bf r}] = \int {\cal L}({\bf r} )d^3{\bf r}dt,
\label{action0}
\end{equation}
in which we have employed a Lagrangian description and ${\bf r}$
represents the coordinates of a fluid element.
The action is a function of the initial coordinate positions ${\bf
  r}_0$ and $t$, and 
 the Lagrangian density for the compressible system is given by 
\begin{equation}
{\cal{L}} =\rho\left(\frac{1}{2} \u^2+ \Omega_0{ ({\bf e}_z}\times {\bf r})\cdot\u 
-\Phi_T -\frac{P}{(\gamma -1)\rho} -\frac {{\bf B}^2}{8\pi} \right) .
\end{equation}
Apart from the second term, arising from the
rotating frame,
this expression is identical to that presented by Ogilvie (2016).
An expression incorporating the second term but without a magnetic field, ${\bf B},$ applicable to
a 2D shearing sheet, has been formulated by Goldreich et al. (1987).
For the incompressible and anelastic cases, the fourth thermal energy term is absent and instead one must
apply a conservation  constraint through the use of a Lagrange
multiplier (see Vasil et al.~2013).
In the incompressible case, the conserved quantity is $\rho$ 
and in the anelastic case it is $\rho/{\overline{ P}}^{1/\gamma}.$

One may suppose that there is a steady state background flow and
 consider perturbations to the particle positions
such that  ${\bf r} \rightarrow {\bf r}+ {\mbox{\boldmath{$\xi$}}},$
where ${\mbox{\boldmath{$\xi$}}}$ is the  Lagrangian displacement, 
related to the Eulerian velocity perturbation via
\begin{equation}
{\bf u}' = 
 \frac{\partial \mbox{\boldmath{$\xi$}}}{\partial t} +{\overline \u}\cdot\nabla\mbox{\boldmath{$\xi$}}.
\end{equation}
When substituted into (\ref{action0}), the action integral is
quadratic in the perturbations, on account of it being stationary in
the background state. 
Thus
the action converts into
\begin{equation}
\mathcal{S}[{\mbox{\boldmath{$\xi$}}}] = 
\int {\cal L}({\mbox{\boldmath{$\xi$}}},\partial {\mbox{\boldmath{$\xi$}}}/\partial t,
\nabla{\mbox{\boldmath{$\xi$}}}) d^3{\bf r}dt,
\label{action1}
\end{equation}
(see Ogilvie 2016, Goldreich et al. 1987 for specific evaluations).
Wave-action conservation follows by noting that the background is independent of $y$
and so expresses translational symmetry in that direction. Noether's theorem then
yields a wave-action conservation law 
\begin{equation}
\frac{\partial (\mrho Q)}{\partial  t} + \nabla\cdot {\bf J}  =0, \label{action}
\end{equation}
(see Goldreich et al. 1987), where the wave-action density is 
\begin{equation}
\mrho Q =\frac{\partial {\cal L}}{ \partial (\partial{\mbox{\boldmath{$\xi$}}}/\partial t)}\cdot\frac{ \partial {\mbox{\boldmath{$\xi$}}}}{\partial y},
\hspace{3mm} \label{Noether1}
\end{equation}
and the wave-action density flux is
\begin{equation}
{\bf J} =\frac{\partial {\cal L}}{ \partial (\partial{\mbox{\boldmath{$\xi$}}}/
\partial {\bf r})}\cdot\frac{ \partial {\mbox{\boldmath{$\xi$}}}}{\partial y} - {\cal L} {\bf e}_y.
\label{Noether2}
\end{equation}
The scalar products in these equations are between the two occurrences of $\mbox{\boldmath{$\xi$}}.$

One may also obtain an energy conservation law for perturbations by
replacing 
${ \partial {\mbox{\boldmath{$\xi$}}}}/{\partial y}$
by ${ \partial {\mbox{\boldmath{$\xi$}}}}/{\partial t}$ in (\ref{Noether1})  and (\ref{Noether2})
and then adding $-{\cal L}$  to  (\ref{Noether1}) while removing the
term $\propto {\bf e}_y$ from (\ref{Noether2}).
This is useful  for discussing
disturbances that do not depend on $y.$ 
However, for the sake of brevity we shall focus  here on the conservation of wave action.

Alternatively, we
can derive conservation laws from the equations
governing the linear perturbations in the Lagrangian formulation
(see eg. Lynden-Bell and Ostriker 1967).
Note that the wave-action conservation law may be rescaled so as to
represent the conservation of $y$-momentum (and then angular momentum)
through an additional multiplication by $R_0$
 (the radial location of the shearing box in the disc). 
This scaling can be determined by  
 incorporating  an  external forcing potential into  the
 linearized equations of motion and considering the consequent injection of energy and momentum
in the modified conservation laws.
In this way $-\rho Q$  can be interpreted as the wave $y$-momentum density.

Wave-action conservation can tell us how the amplitude of
a wave varies as it propagates through a variable background, by
computing the `instantaneous' wave's properties
in the local limit.  
As an illustrative example we consider the anelastic model 
governed by equations    (\ref{ani22}), (\ref{ani3}) and  (\ref{ani4}) 
and  also the compressible model governed by (\ref{comp1}), (\ref{comp2}) and (\ref{comp3}) 
allowing for the incorporation of vertical stratification if needed,  
 all with no cooling. 
Extension to the incompressible case follows by taking the limit $\gamma \rightarrow \infty.$
 
We determine the relevant conservation law for wave action
by first averaging over 
  the $y$ direction. Then
\begin{equation}
Q=\left \langle { \frac{\partial \mbox{\boldmath{$\xi$}}}{\partial t} \cdot 
 \frac{\partial \mbox{\boldmath{$\xi$}}}{\partial y}     +
 {\overline u}_y\frac{\partial \mbox{\boldmath{$\xi$}}}{\partial y} \cdot 
 \frac{\partial \mbox{\boldmath{$\xi$}}}{\partial y}   
      -\Omega_0{\bf e}_z\cdot \left( \frac{\partial \mbox{\boldmath{$\xi$}}}{\partial y} 
      \times{ \mbox{\boldmath{$\xi$}}}
 \right) }\right\rangle
\end{equation}
and  
\begin{equation}
{\bf J}=  \left \langle P' \frac{\partial \mbox{\boldmath{$\xi$}}}{\partial y}\right \rangle ,
\end{equation}
where the angle brackets denote an integral mean over the $y$ domain
under the assumption that periodic boundary conditions apply.
By integrating  (\ref{action}) over the volume of the box $V$, 
given that fresh  wave-action  density  cannot enter through the boundaries,
one discovers that the total   integrated wave-action density,
\begin{equation} \label{connie}
\int_V\mrho Q d^3{\bf r},
\end{equation}
is a constant, which simply expresses conservation 
of the total angular momentum
 associated with the disturbance. It must be stressed that 
when the background's variation is retained in an inconsistent manner,
the correct action conservation cannot be obtained. 

We next apply expression \eqref{connie} to the question of instability.
It is clear that if a small disturbance $\boldsymbol{\xi}$ is to grow (i.e.
 there exists an instability), the  wave-action density
 must change sign somewhere in the domain $V$:  
the only way a disturbance can grow exponentially
somewhere in the box is if there is a cancellation  with an exponentially growing disturbance
with the \emph{opposite} sign of wave action density elsewhere. This is precisely the
 mechanism that drives corotation-type instabilities (Papaloizou \&
 Pringle 1984), which work via the interaction of disturbances
 with opposite signs of angular momentum. To understand these instabilities
 a global treatment is usually required (though see the slender torus
 models of Goldreich et al.~1986, Narayan et al.~1987, and Latter and Balbus 2009).  
 
 On the other hand, Eq.~\eqref{connie} tells us
 highly localised disturbances with wave-action density of a well defined sign 
 cannot grow while they propagate. This precludes instabilities of the
 type explored in the previous subsection.   
 Note that this principle allows velocity amplitudes to increase as the wave enters a low density region,
 but it is incorrect to view this as
 localised exponential growth (and hence instability).
 For perturbations such as these, which vary harmonically in time, the amplitude
of a propagating wave packet  is governed  by $\nabla\cdot{\bf J}=0,$
with a further time average applied to ${\bf J}.$
Its  properties can be determined, apart from a constant amplitude, 
in the local limit for which
variations in the background variables are neglected. 
Then the above equation constrains, and may completely determine, the variation of the amplitude 
as it propagates.

\section{Conclusion}

In this paper we start from a fully global and fully compressible disk
and derive a sequence of local shearing box models
that describe small regions of the disk under various
assumptions. These approximations
are consistent with the original equations via a well defined ordering
scheme and satisfy key conservation laws (energy, potential
vorticity, entropy, etc). We stress that one must be careful in
deriving local models: various terms and background gradients cannot
be thrown in or removed arbitrarily. Spurious instabilities or
other undesirable features may arise.

Slow phenomena on short scales can be described by incompressible or
Boussinesq equations, the two models only differing in the relative
sizes of the fractional density and pressure perturbations. We show
how vertical shear may be incorporated in both, but that only the 
Boussinesq system involves a background gradient in entropy. 

We reiterate the
problems inherent in anelastic models; commonly used
versions of these equations fail to conserve energy. When ad hoc
adjustments are made, conservation is assured only in special cases,
such as isothermality, which precludes the treatment of
convection. A set of conservative anelastic equations is derived that
permits diabaticity, but restricts the characteristic length scale of
phenomena, though not the domain over which the phenomena can range. 
Admittedly, the restriction to small-scales, necessitated by the strong shear in a thin Keplerian disc,
 is a strong imposition,
but it does illustrate the challenges posed by
the anelastic approximation.
Applications of this conservative set include magnetorotational or convective 
turbulence localised to certain layers, or the propagation and
refraction
of wave packets upward in the disc. 

We derive two forms of compressible model, one in which
the characteristic length scales are much less than the scale height 
(the `small compressible shearing box') and one in which they are of
order the scale height (the `large' or vertically stratified box). 
Emphasis is put on the problems that arise when the background
gradients are included in the former, especially the generation of
spurious acoustic overstabilities. We discuss how these relate to the
breaking of wave-action conservation: in particular,
if wave action is properly conserved in the model, localised disturbances with a well defined 
sign for the associated action density do not undergo such overstabilities.

Finally, we point out that in a typical hydrodynamical cascade, 
as energy tumbles to smaller and smaller scales, the flow becomes more
and more incompressible and rotation and shear less and less
important. Ultimately, on some scale above the viscous dissipation
length, the flow may be approximated within a non-shearing
incompressible local model. This should be kept in mind when studying
the microscales in discs, such as the interactions between dust and
turbulent eddies. If the fluid is sufficiently ionised, and a similar
cascade is functioning, then any imposed magnetic field (no matter how weak)
will ultimately dominate on some small scale. In this case, the
flow can be modelled by the equations of reduced MHD (Biskamp 1993),
and indeed recent simulations of the MRI
show evidence of this regime (Zhdankin et al.~2017).

The derivations in this paper were purely hydrodynamic, as the main
issues and problems issue from the thermodynamics, 
but it is straightforward to
generalise these derivations to MHD: the induction equation and
Lorentz force pose no additional complications. On the other hand, 
it is extremely difficult to extend the `locality' of the shearing
box, either by including 
higher order terms arising from the disc's cylindrical geometry
(Pessah and Psaltis 2005), or by relaxing the assumption that $z\lesssim
H_Z$ (McNally and
Pessah 2015). The former case seeds spurious modes on arbitrarily
small-scales, while the latter encounters conservation difficulties,
as pointed out in detail by McNally and Pessah
(2015). In an interesting contrast, the local manifestation of global
phenomena such as warps and eccentricities can be described
consistently in a suitably modified shearing box (Ogilvie \& Latter
2013, Ogilvie \& Barker 2014). 

Though it is possible to generalise the shearing box to
relativistic flow (Heinemann, private communication) one must be
careful with the radial boundary conditions. Shearing periodic
boundaries, as employed in numerical realisations (Riquelme et
al.~2012, Hoshino 2013, 2015), are inconsistent
with Lorentz invariance (see Peters 1983), and produce spurious
effects such as `run-away' particles (Kimura et al.~2016). The
infinite
relativistic shearing sheet, however, may offer a useful platform to
undertake purely theoretical work. 

Finally, as is well known,
two-dimensional razor-thin or vertically integrated
shearing boxes cannot be rigorously derived from three-dimensional
equations on account of the quadratic velocity nonlinearity in the
momentum equation. However, if it is assumed that the planar velocities 
exhibit little to no vertical variation, and that phenomena possesses
planar scales much greater than $H_Z$, it is possible to well-motivate
two-dimensional vertically integrated shearing box equations (e.g.\
Shu and Stewart 1985, Stehle and Spruit 1999): a 2D
incompressible model from the classical anelastic equations, and a 2D compressible
model from the large compressible box.

We finish by stressing the continuing value of the local shearing box 
model in understanding disc phenomena. On account of its simpler geometry,
problems are analytically and numerically easier; they hence permit
researchers to disentangle the salient physical effects and their
relationships, and hence make real progress in our understanding. 
The resolution obtained in numerical simulations is also an advantage
local models wield over global set-ups, where typically one is more
concerned with obtaining a reasonable scale separation between radius
and the vertical scale height, rather than between the input and
dissipation scales in turbulence. In fact, only in local models can
any kind of turbulent inertial
range be simulated adequately; at present `turbulence' in global models
resembles
more a monoscale chaotic flow. 
Of course, local models
have their deficiencies, some of them outlined in Umurhan and Regev
(2008), though we feel most of their criticisms are overstated. 
For instance, problems issuing from symmetries, boundary conditions,
and enhanced fluctuations
can be ameliorated by simply varying the boundary conditions and/or taking
bigger boxes (admittedly certain problems do pose special difficulties
on this count; e.g.\ Fromang et al.~2013). 
Global models also struggle with boundary conditions,
which are often ambiguous, unrealistic, or numerically problematic. 
No model is perfect, and each has its strengths and weaknesses. If we
are alert to these, shearing boxes remain valuable tools in helping us
understand the complicated astrophysical flows around planets, stars,
and black holes.  

\section*{Acknowledgements}

The authors thanks the anonymous reviewer for a set of helpful
comments that improved the presentation of the material, and Tobias
Heinemann for generously reading through a previous draft. 
HNL is partly funded by STFC grant 
ST/L000636/1.

\begin{appendix}

\section{Energy conservation with vertical shear}

The origin of  the energy source term on the right hand of  
(\ref{CONE}) can be understood more generally
by considering the basic equation of motion  (\ref{i2}).
 From this we may obtain
 without  assuming $\nabla\cdot {\bf u} =0,$
\begin{equation}
\d_t(\rho(\tfrac{1}{2} u^2+\Phi)) + 
\nabla\cdot\left[(\tfrac{1}{2}u^2+h+\Phi  )\rho\u
\right]= -\frac{P } {\rho}\frac{D\rho}{Dt} ,\label{CONEF}
\end{equation}
where now we have $h=P/\rho.$
If $\rho$ were constant (\ref{CONEF}) would express conservation of energy.
But in order to be consistent with $q_Z\ne 0,$ this cannot be 
case and the rate of doing ${PdV}$ work provides a source.

Given that $\rho'$ is small, to lowest order in $\epsilon,$ 
 we may set $D\rho/Dt= {\bf u}\cdot\nabla\rho$.  We also replace $P$ by $P - {\overline P}_0$
thus measuring it relative to the constant value in the centre of the box.
In addition, to lowest order in $\epsilon$ we neglect $P'$ and $\rho'.$
 We can then  set $\rho = {\overline \rho}$ and $P ={\overline P} -{\overline P}_0$ and    perform a first order Taylor expansion about the centre of the box
  to obtain the latter  quantity and hence the rate of doing $PdV$ work.
To first order in $\epsilon,$ we obtain
\begin{eqnarray}
&&\d_t(\rho(\tfrac{1}{2} u^2+\Phi)) + 
\nabla\cdot\left[(\tfrac{1}{2}u^2+{ h}+\Phi  )\rho \u
\right]\nonumber\\
&& =-\frac {{\bf u}\cdot (\nabla {\overline \rho})_0}{\rho_0}\left ( (\partial_R {\overline P})_{0}x + (\partial_Z{\overline P})_{0}z \right).\label{CONEF1}
\end{eqnarray}
After some manipulations and use of (\ref{tw}) this can be written as
\begin{eqnarray}
&&\d_t(\rho(\tfrac{1}{2}\rho u^2+\Phi)) + 
\nabla\cdot\left[(\tfrac{1}{2}u^2+{ h}+\Phi +{\cal Q} )\rho \u\right]\nonumber\\
&&= -2{\overline \rho_0}\Omega_0^2q_Z z\,u_x + {\cal Q}  \rho \nabla\cdot \u
\label{CONEF2}
\end{eqnarray}
where
\begin{equation}
{\cal Q}= \frac{(\partial_R {\overline \rho})_{0}(\partial_R {\overline P})_{0}x^2 + (\partial_Z {\overline \rho})_{0}(\partial_Z{\overline P})_{0}z^2+
2(\partial_R {\overline \rho})_{0}(\partial_Z{\overline P})_{0}zx}{2\rho{\overline \rho}_0}.
\end{equation}
When $\nabla\cdot{\bf u}$ is neglected,  the source terms in
(\ref{CONE}) and (\ref{CONEF2}) are seen to be identical.

\section{Potential vorticity conservation in general Boussinesq systems}

In Section 3 we derived conservation laws in the more
straightforward case of vertical stratification. Now we examine the general
barotropic case, for which  $N_R^2H_R^2=N_Z^2H_Z^2$ and $q_Z=0.$ 
We introduce the convenient `total' buoyancy variable
$\theta_s= H_Z\rho'/\mrho_0 + z +xH_Z/H_R$, which transforms the entropy
equation into 
$D\theta_s/Dt=0.$ 
Next
 we take the
curl of \eqref{boozemomdim} 
and obtain an equation for the vorticity in the shearing
box:
\begin{equation}
\frac{D\omi}{Dt} = \omi\cdot\nabla\u - N^2_Z\nabla\theta_s
\times (\ez+\ex H_Z/H_R) \,. \label{eq4a}
\end{equation}
We now  take the inner product of \eqref{eq4a} and
$\nabla\theta_s$, which gives
\begin{equation}
\nabla\theta_s\cdot \frac{D\omi}{Dt} 
= \nabla\theta_s\cdot\left[(\omi\cdot\nabla)\u\right]. \label{eq5a}
\end{equation}
The gradient of the entropy equation is
\begin{equation}
\frac{D\nabla\theta_s}{Dt} = -(\nabla\u)\cdot\nabla\theta_s. \label{eq6a}
\end{equation}
Finally, we examine the total derivative
of $\nabla\theta_s\cdot\omi$, and see that
\begin{align}
\frac{D(\nabla\theta_s\cdot\omi)}{Dt} &= 0
\end{align}
The final result is:
\begin{equation}
\frac{\d \Theta}{\d t} + \nabla\cdot(\Theta\,\u) =0, \label{con2a}
\end{equation}
where the conserved quantity is 
$\Theta=\nabla\theta_s\cdot\omi$. This we regard as the potential vorticity in
the Boussinesq shearing box for a barotropic flow  with general stratification.

\end{appendix}

\end{document}